\documentclass[12pt,english]{article}
\usepackage{lmodern}
\usepackage[T1]{fontenc}
\usepackage[latin9]{inputenc}
\usepackage{geometry}
\usepackage{amsmath}
\usepackage{amssymb}
\usepackage{graphicx}
\usepackage{esint}
\usepackage{subscript}

\makeatletter
\usepackage{color}
\usepackage{hyperref}
\usepackage{cite}
\usepackage{babel}
\usepackage{mathdots}
\usepackage{amsfonts}
\usepackage{mathrsfs}
\usepackage{amsmath}
\usepackage{amssymb}
\usepackage{esint}
\usepackage{graphicx}
\usepackage{youngtab}
\usepackage{multicol}
\usepackage{multirow}
\usepackage{slashed}

\usepackage[titletoc]{appendix}

\allowdisplaybreaks

\numberwithin{equation}{section}

\date{}

\usepackage[usenames,dvipsnames,svgnames,table]{xcolor}
\hypersetup{colorlinks=true, citecolor=red, linkcolor=blue, urlcolor=violet}

\usepackage{titlesec}
\titleformat{\section}{\large\bf}{\thesection}{1em}{}
\titleformat{\subsection}{\normalsize\bf}{\thesubsection}{1em}{}

\makeatother

\usepackage{babel}
\begin{document}

\title{\textbf{\Large{Holographic Spontaneous Parity Breaking and \\ Emergent
Hall Viscosity and Angular Momentum \\\bigskip{}\bigskip{}}}}

\author{\textbf{\normalsize{Dam Thanh Son and Chaolun Wu}}}

\maketitle
\vspace{-16pt}

\begin{center}
\textit{Kadanoff Center for Theoretical Physics and Enrico Fermi Institute}
\par\end{center}

\begin{center}
\vspace{-35pt}

\par\end{center}

\begin{center}
\textit{University of Chicago, Chicago, Illinois 60637, USA}
\par\end{center}

\vspace{-10pt}

\begin{center}
\textit{\small{Email: }}\texttt{\small{dtson@uchicago.edu}}\textit{\small{,
}}\texttt{\small{chaolunwu@uchicago.edu}}
\par\end{center}{\small \par}

\begin{center}

\par\end{center}

\begin{center}
\thispagestyle{empty}
\par\end{center}
\begin{abstract}
We study the spontaneous parity breaking and generating of Hall viscosity
and angular momentum in holographic p+ip model, which can describe
strongly-coupled chiral superfluid states in many quantum systems.
The dual gravity theory, an SU(2) gauge field minimally coupled to
Einstein gravity, is parity-invariant but allows a black hole solution
with vector hair corresponding to a parity-broken superfluid state.
We show that this state possesses a non-vanishing parity-odd transport
coefficient -- Hall viscosity -- and an angular momentum density.
We first develop an analytic method to solve this model near the critical
regime and to take back-reactions into account. Then we solve the
equation for the tensor mode fluctuations and obtain the expression
for Hall viscosity via Kubo formula. We also show that a non-vanishing
angular momentum density can be obtained through the vector mode fluctuations
and the corresponding boundary action. We give analytic results of
both Hall viscosity and angular momentum density near the critical
regime in terms of physical parameters. The near-critical behavior
of Hall viscosity is different from that obtained from a gravitational
Chern-Simons model. We find that the magnitude of Hall viscosity to
angular momentum density ratio is numerically consistent with being
equal to 1/2 at large SU(2) coupling corresponding to the probe limit,
in agreement with previous results obtained for various quantum fluid
systems and from effective theory approaches. In addition, we find
the shear viscosity to entropy density ratio remains above the universal
bound.
\end{abstract}
\newpage{}

{\hypersetup{linkcolor=black}

\tableofcontents{}

}

\bigskip{}

\bigskip{}
\bigskip{}

\section{Introduction}

Systems with broken parity and time-reversal symmetries have long
been attractive and active fields to both experimentalists and theorists
in physics. When these discrete symmetries are allowed to be broken,
additional transport coefficients can arise in the hydrodynamic description
of the systems. In $2+1$-dimensional systems, Hall conductivity,
the parity-odd and dissipationless counterpart of the ordinary longitudinal
conductivity, is probably the most famous and best-studied example.
It is also been known that the viscosity can have a parity-odd and
dissipationless part as well, called Hall viscosity. On contrary to
the parity-even and dissipative shear viscosity, which, in classical
picture of fluids, tends to accelerate or decelerate nearby flows
in the presence of a gradient of velocity, Hall viscosity tends to
repel or attract the nearby flows. The underlying force (sometimes
referred to as ``Lorentz shear force'' in the literature) is perpendicular
to the flow, thus is dissipationless. A pictorial illustration can
be found in \cite{Hughes:2012vg}. An example of Hall viscosity in
classical fluid systems is given by a plasma moving in magnetic field
\cite{Lifshitz:Book}. Recently, Hall viscosity was studied for various
non-relativistic quantum systems such as quantum Hall fluids \cite{Avron:1995fg,Avron:1998,Tokatly:1,Tokatly:2,Read:2008rn,Haldane:2009ke,Read:2011}
and chiral superfluids and superconductors \cite{Read:2008rn,Read:2011},
and relativistic quantum systems such as topological insulators with
massive Dirac fermions \cite{Hughes:2011hv,Hughes:2012vg}. It was
also studied using general approaches such as linear response theory
\cite{Bradlyn:2012ea}, effective field theories \cite{Nicolis:2011ey,Hoyos:2011ez,Hoyos:2013eha,Son:2013rqa},
viscoelastic-electromagnetism \cite{Hidaka:2012rj} and quantum hydrodynamics
of vortex flow \cite{Wiegmann:1211,Wiegmann:1305,Wiegmann:1309}.
Hall viscosity possesses many interesting properties. In quantum Hall
fluids it is related to the Berry curvature and the Wen-Zee shift
\cite{Wen:1992ej}, thus reflects the topological feature of the quantum
Hall states. Hall viscosity also enters as a finite wave number correction
to Hall conductivity \cite{Hoyos:2011ez}. Of particular interest
is a general relation between Hall viscosity $\eta_{H}$ and angular
momentum density $\ell$ of the system:
\begin{equation}
\eta_{H}=-\frac{1}{2}\ell\:,\label{HL_Relation}
\end{equation}
which is derived first in \cite{Read:2008rn} for quantum Hall states
and $p_{x}+ip_{y}$ superfluids, then in \cite{Nicolis:2011ey,Hoyos:2013eha,Son:2013rqa}
from effective field theory methods. In this paper we try to understand
Hall viscosity and the above relation to angular momentum density
in strongly-interacting quantum many-body systems, particularly the
$p_{x}+ip_{y}$ paired states, from the holographic point of view.

Over the last decade, holography, or gauge/gravity duality \cite{Maldacena:1997re,Gubser:1998bc,Witten:1998qj}
has been widely applied to study many strongly interacting systems.
One of its remarkable early successes is to study hydrodynamic transport
coefficients of strongly coupled relativistic conformal fluids \cite{Policastro:2002se,Policastro:2002tn},
in particular the shear viscosity to entropy density ratio \cite{Kovtun:2003wp}
(for an recent review, see \cite{Cremonini:2011iq}), whose theoretical
value obtained via holography is very close to that of the quark-gluon
plasma extracted from RHIC and LHC data. 

Another recent application of holography is on superconducting and
superfluid phase transitions in condensed matter systems. The hope
is to gain insight into systems that cannot be described by the BCS
theory, like high $T_{c}$ superconductors. The superconducting phase
transition is characterized by a charged operator $\mathcal{O}$,
whose expectation value $\langle\mathcal{O}\rangle$ is zero above
a certain critical temperature $T_{c}$ (the normal phase), but becomes
non-zero below $T_{c}$ (the superconducting phase). In the dual gravity
theory, the normal phase at non-zero temperature is usually described
by a charged AdS black hole. The operator $\mathcal{O}$ is dual to
a certain charged matter field $\phi$ that couples to this black
hole. Below the critical temperature $T_{c}$, $\phi$ can develop
a non-trivial profile outside the black hole horizon \cite{Gubser:2005ih,Gubser:2008px}.
The resulting hairy black hole, which is thermodynamically preferred
over the hairless one, describes the superconducting phase. Depending
on whether the matter field $\phi$ is a scalar, a non-Abelian gauge
field, or a symmetric tensor field, it describes holographically the
$s$-wave \cite{Hartnoll:2008vx,Hartnoll:2008kx,Horowitz:2008bn,Herzog:2008he},
$p$-wave \cite{Gubser:2008zu,Gubser:2008wv,Roberts:2008ns} or $d$-wave
\cite{Chen:2010mk,Benini:2010pr} superconductors respectively. In
\cite{Hartnoll:2008kx} it is shown that such holographic superconductors
are Type II superconductors. For reviews in this subject, see \cite{Herzog:2009xv,Horowitz:2010gk}.

A third application of holography is to strongly coupled systems with
broken parity and time-reversal symmetries, such as quantum Hall systems
\cite{HallReview1,HallReview2,HallReview3,HallReview4,HallReview5}.
For $2+1$-dimensional systems, dyonic AdS black hole is a simple
holographic realization of the classical Hall effect and produces
unquantized Hall conductivity \cite{Hartnoll:2007ai}. To address
quantum Hall effects, including the integer and fractional quantized
Hall conductivities, quantum plateau transitions and edge states,
models with various matter fields or brane configurations were considered
\cite{KeskiVakkuri:2008eb,Davis:2008nv,Fujita:2009kw,Bergman:2010gm,Gubankova:2010rc,Kristjansen:2012ny},
and these added structures usually include Chern-Simons terms, which
break the parity of the theories explicitly or spontaneously. In $3+1$
dimensions, two other parity-violating effects -- the chiral magnetic
effect and chiral vortical effect -- are also realized in holographic
models \cite{Erdmenger:2008rm,Banerjee:2008th,Gynther:2010ed,Kalaydzhyan:2011vx,Amado:2011zx,Landsteiner:2011cp,Landsteiner:2011iq,Landsteiner:2012dm,Landsteiner:2012kd}.

Hall viscosity was obtained in \cite{Saremi:2011ab} for the first
time in a holographic model, with a dynamical axion coupled to Chern-Simons
modified gravity \cite{Alexander:2009tp}, and numerics was soon followed
\cite{Chen:2011fs,Chen:2012ti}. \cite{Cai:2012mg,Zou:2013fua} studied
both Hall viscosity and Curl viscosity using similar holographic models
with Chern-Simons terms. Angular momentum generated in holographic
models was also studied \cite{Liu:2012zm,Liu:2013cha}. However, whether
a holographic model can generate both Hall viscosity and angular momentum
density simultaneously and whether their relation (\ref{HL_Relation})
can hold remain mysterious. For example, for the models considered
in both \cite{Jensen:2011xb} and \cite{Liu:2012zm} there exists
an angular momentum but no Hall viscosity. 

What we study in this paper is an overlap of all the aforementioned
areas in holography. We will show that for the holographic $p_{x}+ip_{y}$
model of \cite{Gubser:2008zu}, in the superconducting phase, both
non-vanishing Hall viscosity and angular momentum density emerge.
Using analytic method to compute both of them near the critical regime,
we find that the relation (\ref{HL_Relation}) holds at the probe
limit regime, but has a deviation when back-reactions are taken into
account. This model is different from most other holographic models
constructed for Hall effects and those used in \cite{Saremi:2011ab,Liu:2012zm}
to compute Hall viscosity and angular momentum: it does not contain
an explicit Chern-Simons term in the action, nor external magnetic
field or rotation \cite{Sonner:2009fk}, which all break parity and
time-reversal symmetry in a manifest way. It is known that in some
types of superconducting phase transitions the breaking of $U(1)$
symmetry is accompanied by a spontaneous parity breaking. In the field
theory picture of this model, the parity is broken spontaneously below
the critical temperature by the formation of $p_{x}+ip_{y}$ paired
ground state of the BCS theory. The total orbital angular momentum
of the Cooper pairs is at eigenstate $|lm\rangle=|11\rangle$, which
breaks parity and time-reversal symmetry. In the dual gravity theory
which involves Einstein gravity and $SU(2)$ gauge field, the parity
is broken by the $SU(2)$ hair of the black hole, which is dual to
the $p_{x}+ip_{y}$ paired ground state. This particular background
locks the $\mathbb{Z}_{2}$ symmetry of spatial reflection in the
two conformally flat spatial directions to the $\mathbb{Z}_{2}$ symmetry
in the $SU(2)$ vector space. The $SU(2)$ gauge connection term breaks
the latter $\mathbb{Z}_{2}$ symmetry explicitly, and this breaking
is propagated to the former spatial $\mathbb{Z}_{2}$ symmetry through
the background. This finally produces non-trivial parity-breaking
effects such as the emergence of Hall conductivity and angular momentum
density, and as expected, they are both proportional to the $SU(2)$
gauge coupling. The $p_{x}+ip_{y}$ model has a gapped zero temperature
ground state, which also ensures that the dissipationless transport
can take place and Hall viscosity is non-vanishing. The gap energy
and thermal Hall conductivity was numerically calculated in \cite{Roberts:2008ns}.

The holographic $p_{x}+ip_{y}$ model of \cite{Gubser:2008zu} was
previously studied mostly in the context of superconductivity and
superfluidity. However it is worth to note here that it may have richer
physics yet to explore, for example, the parity-breaking effects on
transport studied in \cite{Roberts:2008ns} and this paper. In fact
the $p_{x}+ip_{y}$ model is more than just a description of chiral
superfluid states in, for example, the A-phase of Helium-3 \cite{StoneRoy:2003,Sauls:2011,Tsutsumi:2012us}
and layered Sr\textsubscript{2}RuO\textsubscript{4} superconductors
\cite{Sr2RuO4-1,Sr2RuO4-2}. It also plays an important role in understanding
the $\nu=5/2$ quantum Hall state and all interesting physics associated
with it, such as non-Abelian anyons and its potential application
to quantum computation (for recent reviews on this subject, see \cite{Stern:Review,Nayak:2008zza,Willet:Review}).

It is worth noting here that \cite{Gubser:2008wv} shows the $p_{x}+ip_{y}$
superconducting state can be unstable upon perturbations and tunnel
to an anisotropic $p$-wave ground state. However, this problem might
be overcome by adding non-linear terms in the action to stabilize
the $p_{x}+ip_{y}$ solution, similar as in \cite{Radu:2011ip}. As
long as the solution remains homogeneous and isotropic, the parity
breaking properties studied in this paper will still hold qualitatively
in the new model, with added corrections from the non-linear effect.

The paper is organized as following. In Section 2, we briefly review
first order relativistic parity-violating hydrodynamics, including
the definition of Hall viscosity in this context and Kubo formulae
associated to it. In Section 3 we give the general formalism of Einstein-$SU(2)$
system, which is the basis where the holographic $p$-wave superconductor
models are built on. In Section 4 we review the $p_{x}+ip_{y}$ model
of \cite{Gubser:2008zu} and propose our refined analytic method to
solve this model near the critical regime. Our method takes the back-reactions
between the metric and the matter field into full consideration. In
the next two sections we compute Hall viscosity and angular momentum
density from tensor and vector mode bulk fluctuations respectively,
and then the ratio between them. In Section 7 the low temperature
limit of the model is investigated. In the last section we will make
conclusion remarks and comments. Except that in Section 3 that we
work in general $d+1$ dimensions, we always work in $3+1$ dimensions
in the gravity theory, thus the dual field theory or condensed matter
systems are in $2+1$ dimensions.

\bigskip{}

\section{Relativistic First Order Parity-Violating Hydrodynamics}

Hydrodynamics is a large-scale effective description of fluids and
many other classical and quantum systems at non-zero temperature.
The fundamental EOMs are the conservation of the energy-stress tensor
and the current: 
\begin{eqnarray}
\nabla_{\mu}T^{\mu\nu} & = & F^{\nu\mu}J_{\mu}\:,\label{eq:Conserv_StressTensor}\\
\nabla_{\mu}J^{\mu} & = & 0\:.\label{eq:Conserv_Current}
\end{eqnarray}
Here we allow the current to couple to an external gauge field whose
strength is $F_{\mu\nu}$. For simplicity we assume there is only
a $U(1)$ symmetry associated with the conserved current. This is
obviously not true like in the case of $SU(2)$ gauge symmetry considered
in the rest of this paper. In that case it is straightforward to generalize
by letting the current and transport coefficients associated with
it carry $SU(2)$ vector indices, but the Kubo formulae for viscosities
will remain the same. When the system has conformal symmetry, there
is an additional equation of state due to scale invariance: 
\begin{equation}
T_{\mu}^{\mu}=0\:.
\end{equation}
This is the case we will study in this paper. To solve the above equations
for a particular system, one need to supplement them with constitutive
relations which specify the form of $T^{\mu\nu}$ and $J^{\mu}$ in
terms of derivative expansion of local macroscopic functions such
as energy density, pressure and velocity field, among others. Terms
allowed in these constitutive relations can be determined based on
symmetries of the systems and thermodynamical considerations, up to
some arbitrary constants to be determined by the underlying microscopic
theory. These constants are the transport coefficients. For relativistic
conformal systems, there is only one possible first order term allowed
by symmetries in each of $T^{\mu\nu}$ and $J^{\mu}$, whose coefficients
are the shear viscosity and conductivity, respectively. 

When parity is not respected, there are additional terms allowed in
the constitutive relations, with additional transport coefficients.
Based on symmetries and thermodynamic considerations,\cite{Jensen:2011xb}
systematically studied this case for relativistic fluid in 2+1 dimensions
and obtained complete first order constitutive relations (the non-relativistic
version was also studied recently in \cite{Kaminski:2013gca}). In
this paper, we are interested in the sourceless case when external
$F^{\mu\nu}=0$, and for simplicity we also assume that the temperature
$T$ and chemical potential $\mu$ are not local functions. The constitutive
relations up to first order in derivatives are 
\begin{eqnarray}
T^{\mu\nu} & = & \varepsilon u^{\mu}u^{\nu}+\left(p-\zeta\nabla_{\alpha}u^{\alpha}-\zeta_{H}\Omega\right)\Delta^{\mu\nu}-\eta\sigma^{\mu\nu}-\eta_{H}\tilde{\sigma}^{\mu\nu}\:,\\
J^{\mu} & = & \rho u^{\mu}\:.
\end{eqnarray}
The velocity field is normalized to $u^{\mu}u_{\mu}=-1$ and 
\begin{eqnarray}
\Delta^{\mu\nu} & = & g^{\mu\nu}+u^{\mu}u^{\nu}\:,\\
\sigma^{\mu\nu} & = & \Delta^{\mu\alpha}\Delta^{\nu\beta}\left(\nabla_{\alpha}u_{\beta}+\nabla_{\beta}u_{\alpha}-g_{\alpha\beta}\nabla_{\gamma}u^{\gamma}\right)\:,\\
\tilde{\sigma}^{\mu\nu} & = & \frac{1}{2}\left(\epsilon^{\mu\alpha\beta}u_{\alpha}\sigma_{\beta}^{\phantom{\beta}\nu}+\epsilon^{\nu\alpha\beta}u_{\alpha}\sigma_{\beta}^{\phantom{\beta}\mu}\right)\:,\\
\Omega & = & -\epsilon^{\mu\nu\alpha}u_{\mu}\nabla_{\nu}u_{\alpha}\:.
\end{eqnarray}
In the above definitions, the expression for shear flow $\sigma^{\mu\nu}$
quoted here is only valid in $2+1$ dimensions. For general $d$-dimensional
spacetime, the last term will have a coefficient $-\frac{2}{d-1}$
rather than $-1$. The definitions of $\tilde{\sigma}^{\mu\nu}$ and
$\Omega$ are only possible for $2+1$ dimensions because the rank-$3$
totally anti-symmetric tensor $\epsilon^{\mu\nu\alpha}$ only exists
in this case. An analogous case in $d=4$ has also been studied, for
example, in \cite{Amado:2011zx,Landsteiner:2012kd}. The coefficients
$\zeta$, $\eta$ and $\eta_{H}$ are bulk, shear and Hall viscosities
and $\varepsilon$, $p$ and $\rho$ are energy density, pressure
and charge density of the system. For a conformal system, $\nabla_{\alpha}u^{\alpha}$
and $\Omega$ parts will drop off, so $\zeta=\zeta_{H}=0$, and this
is the case we will consider in this paper. A double perturbative
expansion in derivatives and metric fluctuations \cite{Saremi:2011ab}
gives 
\begin{equation}
T^{xy}=-ph_{xy}-\eta\frac{\partial}{\partial t}h_{xy}+\frac{1}{2}\eta_{H}\frac{\partial}{\partial t}\left(h_{xx}-h_{yy}\right)+O\left(\vec{\partial},h^{2}\right)\:,\label{eq:Txy_hydroexp}
\end{equation}
where $h_{\mu\nu}$ is the metric fluctuation around the flat Minkowskian
background metric and the coordinates are $x^{\mu}=(t,x,y)$. Using
\begin{equation}
\langle T^{\mu\nu}(x)\rangle_{h}=\langle T^{\mu\nu}(x)\rangle_{h=0}-\frac{1}{2}\int d^{3}x'G_{\textrm{ra}}^{\mu\nu,\alpha\beta}(x,x')h_{\alpha\beta}(x')+O\left(h^{2}\right)\:,
\end{equation}
where the causal 2-point functions of energy-stress tensor in position
space are defined as 
\begin{equation}
G_{\textrm{ra}}^{\mu\nu,\alpha\beta}(x,x')=-i\theta(t-t')\langle[T^{\mu\nu}(x),T^{\alpha\beta}(x')]\rangle
\end{equation}
and those in momentum space are defined as 
\begin{equation}
G_{\textrm{ra}}^{\mu\nu,\alpha\beta}(k)=\int d^{3}xe^{-ikx}G_{ra}^{\mu\nu,\alpha\beta}(x,0)
\end{equation}
with momentum $k^{\mu}=(\omega,\vec{k})$, we obtain the hydrodynamic
expansions for the following 2-point functions 
\begin{eqnarray}
G_{\textrm{ra}}^{xy,xx-yy}(\omega,\vec{k}=0) & = & 2i\eta_{H}\omega+O\left(\omega^{2}\right)\:,\label{eq:Kubo_HallVisc}\\
G_{\textrm{ra}}^{xy,xy}(\omega,\vec{k}=0) & = & p-i\eta\omega+O\left(\omega^{2}\right)\:.\label{eq:Kubo_ShearVisc}
\end{eqnarray}
These will then give the Kubo formulae for the viscosities. 

At this point it is reasonable to ask whether the above formulae are
valid and can be applied to the calculation of holographic $p_{x}+ip_{y}$
model. There are two subtleties. The first one is regarding the global
symmetry. Clearly, the above analysis and that of \cite{Jensen:2011xb}
assume only a global $U(1)$ gauge symmetry, but the model to be discussed
in the rest of this paper has an $SU(2)$ global symmetry. Some terms
in the above equations, particularly those involving electromagnetic
response in the conservation equations and hydrodynamic expansion,
will change once the global gauge group is changed. For example, both
the current $J^{\mu}$ and field strength $F_{\mu\nu}$ will be $SU(2)$-valued
now. But the part involving gravitational response, such as $u^{\mu}$,
$\sigma^{\mu\nu}$ and $\tilde{\sigma}^{\mu\nu}$, will not change.
The only assumption lying behind (\ref{eq:Txy_hydroexp}), (\ref{eq:Kubo_HallVisc})
and (\ref{eq:Kubo_ShearVisc}) are homogeneity and isotropy, not the
global gauge symmetry. A further subtlety is that the ground state
of holographic $p_{x}+ip_{y}$ model, (\ref{BackgroundAnsatz}), seems
to break the apparent rotational invariance and spoil isotropy. As
explained in the text below (\ref{BackgroundAnsatz}), this apparent
``breaking'' of spatial rotational symmetry is restored by combining
the rotational symmetry of the $SU(2)$ gauge group. Since the energy-stress
tensor is an $SU(2)$ singlet which does not see the rotation in the
gauge group, the isotropy is preserved in the hydrodynamic analysis
involving only the energy-stress tensor. Thus the formulae (\ref{eq:Kubo_HallVisc})
and (\ref{eq:Kubo_ShearVisc}) are still valid. 

The second subtlety is about the extra degrees of freedom in superfluids,
namely the superfluid velocity $\xi_{\mu}=\partial_{\mu}\varphi-A_{\mu}$,
where $\varphi$ is the Goldstone boson. On contrary, the velocity
$u^{\mu}$ we introduce before is the normal fluid velocity. In general,
they are both non-vanishing and point on different directions in the
lab frame, thus there are additional first order derivative terms
involving $\xi^{\mu}$ that can be added to constitutive relations
of $T^{\mu\nu}$ and $J^{\mu}$ and give rise to new transport coefficients
and possibly modify the existing Kubo formulae as well. Superfluid
hydrodynamics in 3+1 dimensions has been studied in \cite{Bhattacharya:2011eea,Bhattacharya:2011tra,Neiman:2011mj}
and that in d+1 dimensions with Lifshitz scaling recently in \cite{Chapman:2014hja}
and in \cite{Hoyos:2014nua} for 2+1-dimensional non-Abelian case.
\cite{Hoyos:2014nua} shows that for the case relevant to ours, the
$SU(2)$ superfluids, the Kubo formula for Hall viscosity (\ref{eq:Kubo_HallVisc})
remains valid. However, \cite{Hoyos:2014nua} also shows that there
are additional first order transport coefficients due to superfluid
velocity, i.e. what they call $\tilde{\eta}_{H}$, the ``locking
dependent Hall viscosity'', and $\kappa_{H}$, whose Kubo formulae
are given by 2-point functions of energy-stress tensor and $SU(2)$
current. The $\tilde{\eta}_{H}$ is different from the Hall viscosity
$\eta_{H}$ that is studied in the rest of this paper and in the previous
literature, thus will not be further considered in this paper, even
though this quantity itself is interesting on its own and deserves
further study. We will concentrate on the Hall viscosity given by
(\ref{eq:Kubo_HallVisc}).

The holographic prescription for computing causal 2-point functions
had been studied in \cite{Herzog:2002pc} and that for higher n-point
functions in \cite{Barnes:2010jp,Arnold:2011ja}. In the rest of this
paper we will follow those prescriptions to compute the above two
2-point functions for holographic $p_{x}+ip_{y}$ superconductor model
of \cite{Gubser:2008zu} and obtain the viscosities in that model.

\bigskip{}

\section{Einstein-$SU(2)$ System}

\subsection{Bulk and Boundary Actions}

In this section, for generality we will work in $(d+1)$-dimensional
curved spacetime. $z$ is the radial coordinate and $z=\infty$ is
where the $d$-dimensional time-like boundary locates. The bulk action
for Einstein-$SU(2)$ system is

\begin{equation}
S_{\textrm{bulk}}=\frac{1}{2\kappa^{2}}\int d^{d+1}x\sqrt{-g}\left\{ \mathcal{R}-2\Lambda-\frac{1}{4}\left(F_{\mu\nu}^{\mathbf{I}}\right)^{2}\right\} \:,
\end{equation}
where the cosmological constant $\Lambda=-\frac{d(d-1)}{2R^{2}}$
and $R$ the AdS radius. The $SU(2)$ field strength is 

\begin{equation}
F_{\mu\nu}^{\mathbf{I}}=\partial_{\mu}A_{\nu}^{\mathbf{I}}-\partial_{\nu}A_{\mu}^{\mathbf{I}}+\lambda\epsilon^{\mathbf{IJK}}A_{\mu}^{\mathbf{J}}A_{\nu}^{\mathbf{K}}\:,
\end{equation}
where $\lambda$ is the Yang-Mills coupling, $\mathbf{I},\mathbf{J},\mathbf{K}=\mathbf{1},\mathbf{2},\mathbf{3}$
and $\epsilon^{\mathbf{IJK}}$ is the totally antisymmetric tensor
with $\epsilon^{\mathbf{123}}=1$. The boundary terms include the
Gibbons-Hawking term

\begin{equation}
S_{\textrm{GH}}=\frac{1}{\kappa^{2}}\int_{z=\infty}d^{d}x\sqrt{-\gamma}K
\end{equation}
and a counter term

\begin{equation}
S_{\textrm{ct}}=-\frac{d-1}{\kappa^{2}R}\int_{z=\infty}d^{d}x\sqrt{-\gamma}\:,
\end{equation}
where $\hat{n}_{\mu}$ is the outgoing unit normal 1-form of the boundary,
$\gamma_{\mu\nu}=g_{\mu\nu}-\hat{n}_{\mu}\hat{n}_{\nu}$ is the induced
metric on the boundary and $K=\nabla_{\mu}\hat{n}^{\mu}$ is the extrinsic
curvature of the boundary. To compute first order hydrodynamics this
single counter term is enough. For higher order hydrodynamics, one
need to include more counter terms such as the boundary Ricci scalar
etc \cite{Emparan:1999pm,Kraus:1999di,deHaro:2000xn}.

\subsection{Perturbative Expansion of Actions and EOMs}

To compute 2-point functions, we perturbatively expand the on-shell
actions around the background up to second order in field fluctuations.
The metric and gauge fields are 
\begin{eqnarray}
g_{\mu\nu} & = & \bar{g}_{\mu\nu}+h_{\mu\nu}\:,\\
A_{\mu}^{\mathbf{I}} & = & \bar{A}_{\mu}^{\mathbf{I}}+a_{\mu}^{\mathbf{I}}\:,
\end{eqnarray}
where $\bar{g}_{\mu\nu}$ and $\bar{A}_{\mu}^{\mathbf{I}}$ are the
background and $h_{\mu\nu}$ and $a_{\mu}^{\mathbf{I}}$ are fluctuations.
To fully consider the back-reactions of the gauge fields on the metric,
we assume $h_{\mu\nu}$ and $a_{\mu}^{\mathbf{I}}$ are of the same
order. The first order on-shell action which is linear in fluctuations
is 
\begin{equation}
S_{\textrm{bulk}}^{(1)}=\frac{1}{2\kappa^{2}}\int d^{d+1}x\partial_{\mu}\left\{ \sqrt{-\bar{g}}\left(\bar{\nabla}_{\nu}h^{\mu\nu}-\bar{\nabla}^{\mu}h-\bar{F}^{\mathbf{I}\mu\nu}a_{\nu}^{\mathbf{I}}\right)\right\} \:.\label{S1_onshell}
\end{equation}
The second order on-shell action quadratic in fluctuations is%
\footnote{In this paper we define the symmetrization $A_{(\mu}B_{\nu)}\equiv A_{\mu}B_{\nu}+A_{\nu}B_{\mu}$
and the anti-symmetrization $A_{[\mu}B_{\nu]}\equiv A_{\mu}B_{\nu}-A_{\nu}B_{\mu}$
without the factor of $\frac{1}{2}$.%
} 
\begin{eqnarray}
S_{\textrm{bulk}}^{(2)} & = & \frac{1}{4\kappa^{2}}\int d^{d+1}x\partial_{\mu}\Big\{\sqrt{-\bar{g}}\Big[\frac{1}{2}h\bar{\nabla}_{\nu}h^{\mu\nu}+\frac{3}{2}h^{\mu\nu}\bar{\nabla}_{\nu}h-h^{\rho\sigma}\bar{\nabla}_{\rho}h_{\sigma}^{\mu}-2h^{\mu\rho}\bar{\nabla}^{\sigma}h_{\rho\sigma}\nonumber \\
 &  & \qquad+\frac{3}{2}h^{\rho\sigma}\bar{\nabla}^{\mu}h_{\rho\sigma}-\frac{1}{2}h\bar{\nabla}^{\mu}h-a_{\nu}^{\mathbf{I}}\left(\frac{1}{2}\bar{F}^{\mathbf{I}\mu\nu}h+\bar{F}_{\phantom{\mathbf{I}}\rho}^{\mathbf{I}\phantom{\rho}[\mu}h^{\nu]\rho}+F^{\mathbf{I}(1)\mu\nu}\right)\Big]\Big\}\:.\label{S2_onshell}
\end{eqnarray}
Here all co-variant derivative $\bar{\nabla}$ and raising and lowering
indices are with respect to the background metric $\bar{g}_{\mu\nu}$,
with $h\equiv h_{\mu}^{\mu}$ and 
\begin{equation}
F_{\mu\nu}^{\mathbf{I}(1)}=\bar{\nabla}_{[\mu}a_{\nu]}^{\mathbf{I}}+\lambda\epsilon^{\mathbf{IJK}}\bar{A}_{[\mu}^{\mathbf{J}}a_{\nu]}^{\mathbf{K}}\:.
\end{equation}
These actions are written as integrals of total derivatives, which
means they are boundary terms. Choosing the gauge condition $\bar{g}_{\mu z}=0$
for $\mu\neq z$ and $h_{\mu z}=0$ for any $\mu$, the first order
boundary actions are 
\begin{eqnarray}
S_{\textrm{GH}}^{(1)} & = & \frac{1}{2\kappa^{2}}\int_{z=\infty}d^{d}x\sqrt{-\bar{\gamma}}\left(\bar{K}+\bar{\hat{n}}^{\mu}\bar{\nabla}_{\mu}\right)h\:,\\
S_{\textrm{ct}}^{(1)} & = & -\frac{d-1}{2\kappa^{2}R}\int_{z=\infty}d^{d}x\sqrt{-\bar{\gamma}}h\:,
\end{eqnarray}
and the second order boundary actions are 
\begin{eqnarray}
S_{\textrm{GH}}^{(2)} & = & \frac{1}{4\kappa^{2}}\int_{z=\infty}d^{d}x\sqrt{-\bar{\gamma}}\left\{ \left(\bar{K}+\bar{\hat{n}}^{\mu}\bar{\nabla}_{\mu}\right)\left(\frac{1}{2}h^{2}-h^{\mu\nu}h_{\mu\nu}\right)\right\} \:,\\
S_{\textrm{ct}}^{(2)} & = & -\frac{d-1}{4\kappa^{2}R}\int_{z=\infty}d^{d}x\sqrt{-\bar{\gamma}}\left\{ \frac{1}{2}h^{2}-h^{\mu\nu}h_{\mu\nu}\right\} \:.
\end{eqnarray}
 The background EOMs are 
\begin{eqnarray}
\bar{\mathcal{R}}_{\mu\nu}-\frac{1}{2}\bar{\mathcal{R}}\bar{g}_{\mu\nu}+\Lambda\bar{g}_{\mu\nu} & = & \frac{1}{2}\left[\bar{F}_{\mu\rho}^{\mathbf{I}}\bar{F}_{\phantom{\mathbf{I}}\nu}^{\mathbf{I}\phantom{\nu}\rho}-\frac{1}{4}\bar{g}_{\mu\nu}\left(\bar{F}^{\mathbf{I}}\right)^{2}\right]\:,\\
\bar{\nabla}_{\mu}\bar{F}^{\mathbf{I}\mu\nu}+\lambda\epsilon^{\mathbf{IJK}}\bar{A}_{\mu}^{\mathbf{J}}\bar{F}^{\mathbf{K}\mu\nu} & = & 0\:,
\end{eqnarray}
 and the linearized EOMs are 
\begin{align}
\left[\bar{\nabla}^{2}+\bar{\mathcal{R}}-2\Lambda-\frac{1}{4}\left(\bar{F}^{\mathbf{I}}\right)^{2}\right]h_{\mu\nu}+\bar{\nabla}_{\mu}\bar{\nabla}_{\nu}h-\bar{\nabla}_{\rho}\bar{\nabla}_{(\mu}h_{\nu)}^{\rho}\qquad\qquad\qquad\qquad\qquad\nonumber \\
=h^{\rho\sigma}\left(\bar{F}_{\mu\rho}^{\mathbf{I}}\bar{F}_{\nu\sigma}^{\mathbf{I}}-\frac{\bar{g}_{\mu\nu}}{d-1}\bar{F}_{\rho\eta}^{\mathbf{I}}\bar{F}_{\phantom{\mathbf{I}}\sigma}^{\mathbf{I}\phantom{\sigma}\eta}\right)+\bar{F}_{\rho(\mu}^{\mathbf{I}}F_{\phantom{\mathbf{I}(1)}\nu)}^{\mathbf{I}(1)\phantom{\nu}\rho}+\frac{1}{d-1}\bar{g}_{\mu\nu}\bar{F}^{\mathbf{I}\rho\sigma}F_{\rho\sigma}^{\mathbf{I}(1)}\:,\label{LinearEinstein}
\end{align}
\begin{align}
\bar{\nabla}_{\mu}\left(F^{\mathbf{I}(1)\mu\nu}-\bar{F}_{\phantom{\mathbf{I}}\rho}^{\mathbf{I}\phantom{\rho}\nu}h^{\mu\rho}\right)+\lambda\epsilon^{\mathbf{IJK}}\bar{A}_{\mu}^{\mathbf{J}}\left(F^{\mathbf{K}(1)\mu\nu}-\bar{F}_{\phantom{\mathbf{K}}\rho}^{\mathbf{K}\phantom{\rho}\nu}h^{\mu\rho}\right)\qquad\qquad\qquad\qquad\qquad\nonumber \\
+\bar{F}_{\phantom{\mathbf{I}}\rho}^{\mathbf{I}\phantom{\rho}\mu}\bar{\nabla}_{\mu}h^{\nu\rho}+\frac{1}{2}\bar{F}^{\mathbf{I}\mu\nu}\bar{\nabla}_{\mu}h+\lambda\epsilon^{\mathbf{IJK}}a_{\mu}^{\mathbf{J}}\bar{F}^{\mathbf{K}\mu\nu}=0\:.\label{LinearYangMills}
\end{align}

\bigskip{}

\section{Holographic $p_{x}+ip_{y}$ Model}

\subsection{Background and its Symmetries}

A general discussion on the AdS-black hole type solutions to the Einstein-$SU(2)$
system can be found in \cite{Manvelyan:2008sv}. Here we will only
restrain to the simple model of \cite{Gubser:2008zu}. We now go back
to $d=3$ case and work in it for the rest of this paper. We choose
the ansatz for the background to be 
\begin{equation}
\left\{ \begin{aligned} & ds^{2}=-F(z)dt^{2}+\frac{1}{F(z)}dz^{2}+r(z)^{2}\left(dx^{2}+dy^{2}\right)\\
 & \bar{A}_{t}^{\mathbf{3}}(z)\equiv\Phi(z),\quad\bar{A}_{x}^{\mathbf{1}}(z)=\bar{A}_{y}^{\mathbf{2}}(z)\equiv A(z)
\end{aligned}
\right.\label{BackgroundAnsatz}
\end{equation}
and all other background gauge fields vanishing. $z=\infty$ is the
boundary and $z=z_{H}$ is the horizon. When $A(z)=0$, the above
background, and thus the ground state of the dual field theory, has
two separate $U(1)$ symmetries, one related to the rotation in $(x,y)$-plane
and the other to the rotation in $(\mathbf{1},\mathbf{2})$-vector
space. The appearance of non-vanishing $A(z)$ breaks both symmetries,
but preserves a combination of them -- the $U(1)$ symmetry of the
joint rotations by the same angle in both $(x,y)$-plane and $(\mathbf{1},\mathbf{2})$-vector
space: 
\[
U(1)_{\theta}^{xy}\otimes U(1)_{\phi}^{\mathbf{12}}\rightarrow U(1)_{\theta=\phi}^{xy,\mathbf{12}}\:.
\]
We can see that by introducing the non-vanishing $A(z)$ background,
the originally separate symmetries in spacetime and $SU(2)$ vector
space are ``locked'' together. Similarly $A(z)$ breaks the separate
parity symmetries in both spaces. When $\lambda=0$, i.e. when the
$SU(2)$ gauge field becomes a product of three $U(1)$ fields, the
joint parity symmetry in both spaces (reflection applied to both $(x,y)$-plane
and $(\mathbf{1},\mathbf{2})$-vector space simultaneously) is preserved:
\[
\mathbb{Z}_{2}^{xy}\otimes\mathbb{Z}_{2}^{\mathbf{12}}\rightarrow\mathbb{Z}_{2}^{xy,\mathbf{12}}\:.
\]
However, when $\lambda\neq0$, the parity in $(\mathbf{1},\mathbf{2})$-vector
space is broken explicitly by the non-Abelian gauge connection $\lambda\epsilon^{\mathbf{IJK}}A_{\mu}^{\mathbf{J}}A_{\nu}^{\mathbf{K}}$
in the field strength, thus the parity in $(x,y)$-plane is also broken
indirectly by this gauge connection term through the ``locking''
mechanism introduced by $A(z)$. Now we have a spacetime parity-breaking
ground state thus the theory ``appears'' to be parity-broken and
will have non-vanishing parity-violating transport coefficients such
as Hall viscosity and Hall conductivity. In summary, to reach a spacetime
parity-violating state, we first introduce a non-vanishing $A(z)$
to lock the spacetime symmetries and $SU(2)$ vector space symmetries
together, then break the $SU(2)$ parity explicitly by making it non-Abelian,
and this breaking will propagate to spatial parity.

The EOMs for background fields are 
\begin{eqnarray}
2r(z)\left(\frac{d^{2}}{dz^{2}}r(z)\right)+\left(\frac{d}{dz}A(z)\right)^{2}+\frac{\lambda^{2}\Phi(z)^{2}}{F(z)^{2}}A(z)^{2} & = & 0\:,\label{EQ1}\\
\left(\frac{d^{2}}{dz^{2}}F(z)\right)-2\frac{F(z)}{r(z)^{2}}\left(\frac{d}{dz}r(z)\right)^{2}-\left(\frac{d}{dz}\Phi(z)\right)^{2}-\frac{\lambda^{2}}{r(z)^{4}}A(z)^{4} & = & 0\:,\label{EQ2}\\
\frac{d}{dz}\left[r(z)^{2}\left(\frac{d}{dz}\Phi(z)\right)\right]-\frac{2\lambda^{2}\Phi(z)}{F(z)}A(z)^{2} & = & 0\:,\label{EQ4}\\
\frac{d}{dz}\left[F(z)\left(\frac{d}{dz}A(z)\right)\right]+\lambda^{2}\left(\frac{\Phi(z)^{2}}{F(z)}-\frac{A(z)^{2}}{r(z)^{2}}\right)A(z) & = & 0\:,\label{EQ3}
\end{eqnarray}
with a constraint equation derived from the trace of Einstein equation
\begin{equation}
\left(\frac{d^{2}}{dz^{2}}F(z)\right)+4\frac{F(z)}{r(z)}\left(\frac{d^{2}}{dz^{2}}r(z)\right)+2\frac{F(z)}{r(z)^{2}}\left(\frac{d}{dz}r(z)\right)^{2}+\frac{4}{r(z)}\left(\frac{d}{dz}r(z)\right)\left(\frac{d}{dz}F(z)\right)=\frac{12}{R^{2}}\:.\label{EQ5}
\end{equation}
 Given that (\ref{EQ1})-(\ref{EQ3}) are solved, (\ref{EQ5}) only
fixes an integration constant (near boundary leading order coefficient
of $F(z)$) in terms of the AdS radius $R$, thus it is not an independent
differential equation, but rather an algebraic equation. This fact
will play a role in later calculations.

\subsection{Boundary Conditions and Thermodynamical Functions}

The boundary conditions are imposed near the boundary $z=\infty$
by requiring the metric goes asymptotic AdS. Solving the above five
equations near the boundary, we get
\begin{equation}
\left\{ \begin{aligned} & r(z)=\frac{z}{R}+r_{1}+O\left(\frac{1}{z^{3}}\right)\\
 & F(z)=\left(\frac{z}{R}+r_{1}\right)^{2}+\frac{\Gamma}{z}+O\left(\frac{1}{z^{2}}\right)\\
 & \Phi(z)=\Phi_{0}+\frac{\Phi_{1}}{z}+O\left(\frac{1}{z^{2}}\right)\\
 & A(z)=\alpha_{0}+\frac{\alpha_{1}}{z}+O\left(\frac{1}{z^{2}}\right)
\end{aligned}
\right.\:,\label{NB Conditions}
\end{equation}
where $r_{1}$, $\Gamma$, $\Phi_{0}$, $\Phi_{1}$, $\alpha_{0}$
and $\alpha_{1}$ are constants. Two of them will be determined by
two physical conditions to be discussed later, and the rest will be
determined by the following (regularity) conditions near the horizon
$z=z_{H}$:
\begin{equation}
\left\{ \begin{aligned} & r(z)=r(z_{H})+O\left(z-z_{H}\right)\\
 & F(z)=4\pi T(z-z_{H})+O\left((z-z_{H})^{2}\right)\\
 & \Phi(z)=O\left(z-z_{H}\right)\\
 & A(z)=A(z_{H})+O\left(z-z_{H}\right)
\end{aligned}
\right.\:,\label{NH Conditions}
\end{equation}
where $T$ is the Hawking temperature of the black hole and equals
to the temperature of the field theory system on the boundary. The
entropy density $s$, energy density $\varepsilon$, chemical potential
$\mu$, charge density $\rho$ and order parameter $\langle\mathcal{O}\rangle$
can be expressed in the above asymptotic constants \cite{Gubser:2008zu}:
\begin{eqnarray}
s & = & \frac{2\pi}{\kappa^{2}}r(z_{H})^{2}\:,\label{EntropyDensity}\\
\varepsilon & = & -\frac{\Gamma}{\kappa^{2}R^{2}}\:,\\
\mu & = & \frac{\Phi_{0}}{2R}\:,\\
\rho & = & -\frac{\Phi_{1}}{\kappa^{2}R}\:,\\
\langle\mathcal{O}\rangle & = & \frac{\alpha_{1}}{\kappa^{2}R}\:,
\end{eqnarray}
 and $\alpha_{0}$ can be identified with an external source $J\sim\alpha_{0}$.
Since we are looking for spontaneous symmetry breaking without an
external source, the first physical (boundary) condition we impose
is the vanishing of the source: 
\begin{equation}
\alpha_{0}=0\:.\label{NoSource}
\end{equation}
 The second physical condition is to fix either $\rho$ or $\mu$,
depending on which ensemble one choose: 
\begin{equation}
\begin{cases}
\rho=\textrm{constant} & \qquad\textrm{(Canonical\,\ Ensemble)}\\
\mu=\textrm{constant} & \qquad\textrm{(Grand\,\ Canonical\,\ Ensemble)}
\end{cases}\:.\label{EnsembleCondition}
\end{equation}
The characteristic function of the Canonical Ensemble -- the Helmholtz
free energy density $f_{\textrm{Helmholtz}}$ and that of the Grand
Canonical Ensemble -- the Grand Potential density $\Omega_{\textrm{Grand}}$
(equal to minus of the pressure) are 
\begin{eqnarray}
f_{\textrm{Helmholtz}} & = & \varepsilon-Ts\:,\label{FreeEnergyDef}\\
\Omega_{\textrm{Grand}} & = & f_{\textrm{Helmholtz}}-\mu\rho\:.\label{GrandPotentialDef}
\end{eqnarray}

\subsection{Background On-Shell Action}

The on-shell background bulk action is 
\begin{equation}
\bar{S}_{\textrm{bulk}}=\frac{1}{2\kappa^{2}}\int d^{3}x\int_{z_{H}}^{\infty}dz\left\{ \sqrt{-\bar{g}}\left[\mathcal{\bar{R}}-2\Lambda-\frac{1}{4}\left(\bar{F}_{\mu\nu}^{\mathbf{I}}\right)^{2}\right]\right\} \:.
\end{equation}
 By adding to the Lagrangian (the integrand inside ``$\left\{ \,\right\} $'')
the following combination of background equations: $\frac{1}{2}r(z)^{2}\cdot\left[\textrm{(\ref{EQ2})}+\textrm{(\ref{EQ5})}\right]-A(z)\cdot\textrm{(\ref{EQ3})}$,
the integrand becomes a total derivative. Noticing the near horizon
conditions (\ref{NH Conditions}), there is no contribution from the
horizon. Thus we have 
\begin{equation}
\bar{S}_{\textrm{bulk}}=\frac{1}{2\kappa^{2}}\int_{z=\infty}d^{3}x\left\{ -F(z)\frac{d}{dz}\left(r(z)^{2}+\frac{1}{2}A(z)^{2}\right)\right\} \:.
\end{equation}
 Together with the on-shell background boundary terms 
\begin{equation}
\bar{S}_{\textrm{GH}}+\bar{S}_{\textrm{ct}}=\frac{1}{2\kappa^{2}}\int_{z=\infty}d^{3}x\left\{ r(z)\left[r(z)\left(\frac{d}{dz}F(z)\right)+4F(z)\left(\frac{d}{dz}r(z)\right)-\frac{4}{R}r(z)^{2}\sqrt{F(z)}\right]\right\} \:,
\end{equation}
 the total on-shell background action is 
\begin{equation}
\bar{S}=\frac{1}{2\kappa^{2}}\int_{z=\infty}d^{3}x\left\{ \frac{d}{dz}\left(r(z)^{2}F(z)\right)-F(z)A(z)\left(\frac{d}{dz}A(z)\right)-\frac{4}{R}r(z)^{2}\sqrt{F(z)}\right\} \:.
\end{equation}
 Applying the boundary condition (\ref{NB Conditions}) it can be
written as 
\begin{equation}
\bar{S}=\frac{1}{2\kappa^{2}}\int_{z=\infty}d^{3}x\frac{-\Gamma+\alpha_{0}\alpha_{1}}{R^{2}}\:.
\end{equation}
 The grand potential density equals to $-T$ multiplying the Euclidean
on-shell action ($t$ is integrated from $0$ to $\frac{1}{T}$) mod
the volume: 
\begin{equation}
\Omega_{\textrm{Grand}}=-\frac{T}{V}\bar{S}_{\textrm{Euclidean}}=\frac{\Gamma-\alpha_{0}\alpha_{1}}{2\kappa^{2}R^{2}}\:,\label{GrandPotentialExpression}
\end{equation}
 and by (\ref{GrandPotentialDef}) the free energy density is 
\begin{equation}
f_{\textrm{Helmholtz}}=\frac{\Gamma-\Phi_{0}\Phi_{1}-\alpha_{0}\alpha_{1}}{2\kappa^{2}R^{2}}\:.\label{FreeEnergyExpression}
\end{equation}

There is a useful identity for the background fields that can directly
link the constants in near-boundary conditions (\ref{NB Conditions})
to those in near-horizon conditions (\ref{NH Conditions}). The combination
of background equations $-F(z)\cdot\textrm{(\ref{EQ1})}+r(z)^{2}\cdot\textrm{(\ref{EQ2})}-\Phi(z)\cdot\textrm{(\ref{EQ4})}-A(z)\cdot\textrm{(\ref{EQ3})}$
is a total derivative, thus its integral is a constant: 
\begin{equation}
r(z)^{2}\frac{d}{dz}\left(F(z)-\frac{1}{2}\Phi(z)^{2}\right)-F(z)\frac{d}{dz}\left(r(z)^{2}+\frac{1}{2}A(z)^{2}\right)=\textrm{constant}\:.
\end{equation}
 Evaluating it at both horizon and boundary using (\ref{NH Conditions})
and (\ref{NB Conditions}), we have 
\begin{equation}
4\pi Tr(z_{H})^{2}=\frac{1}{R^{2}}\left(-3\Gamma+\Phi_{0}\Phi_{1}+\alpha_{0}\alpha_{1}\right)\:.\label{HB Relation}
\end{equation}
 So through (\ref{HB Relation}) one can see that the free energy
density obtained above through the on-shell action and (\ref{GrandPotentialDef})
is indeed the same as one can obtain directly from (\ref{FreeEnergyDef})
by computing its right hand side.

\subsection{AdS-Reissner-Nordström Solution}

One solution to equations (\ref{EQ1})-(\ref{EQ5}) is the AdS-Reissner-Nordström
(AdS-RN) solution: 
\begin{equation}
\left\{ \begin{aligned} & r^{(0)}(z)=\frac{z}{R}\\
 & F^{(0)}(z)=\frac{z^{2}}{R^{2}}-\left(1+\frac{q^{2}}{4\lambda^{2}R^{2}}\right)\frac{z_{H}^{3}}{R^{2}z}+\frac{q^{2}z_{H}^{4}}{4\lambda^{2}R^{4}z^{2}}\\
 & \Phi^{(0)}(z)=\frac{qz_{H}}{\lambda R^{2}}\left(1-\frac{z_{H}}{z}\right)\\
 & A^{(0)}(z)=0
\end{aligned}
\right.\:,\label{AdSRN Solution}
\end{equation}
 where $q$ is the dimensionless charge and it is related to the temperature
$T$ and other parameters as 
\begin{equation}
T=\frac{3z_{H}}{4\pi R^{2}}\left(1-\frac{q^{2}}{12\lambda^{2}R^{2}}\right)\:.\label{Temperature}
\end{equation}
 The parameter $q$ and $T$ are confined between two limiting cases
- the Schwarzschild limit 
\begin{equation}
\left\{ \begin{aligned} & T=\frac{3z_{H}}{4\pi R^{2}}\\
 & q=0
\end{aligned}
\right.\label{Schwarzschild Limit}
\end{equation}
 and the Extremal limit 
\begin{equation}
\left\{ \begin{aligned} & T=0\\
 & q=2\sqrt{3}\lambda R
\end{aligned}
\right.\label{Extremal Limit}
\end{equation}
 and as charge $q$ increases, the temperature $T$ decreases. The
relation between $\mu$, $\rho$ and $T$ given by the AdS-Reissner-Nordström
solution is 
\begin{equation}
\rho=\frac{4\pi T\mu R^{2}}{3\kappa^{2}}\left(1+\sqrt{1+\frac{3\mu^{2}}{4\pi^{2}T^{2}}}\right)\:.\label{AdSRN Thermal Relation}
\end{equation}

\subsection{Analytic Approach to the Symmetry-Breaking Solution}

When the temperature $T$ is below a certain critical temperature
$T_{c}$ there exists another non-trivial solution to (\ref{EQ1})-(\ref{EQ5})
that satisfies boundary conditions (\ref{NB Conditions}), (\ref{NH Conditions}),
(\ref{NoSource}) and (\ref{EnsembleCondition}). This symmetry-breaking
solution have been systematically discussed and numerically computed
in \cite{Gubser:2008zu}. Analytic approaches to solve similar models
near the critical temperature in the Probe limit have also been studied
in \cite{Siopsis:2010uq,Zeng:2010zn}. The Probe limit is where the
Yang-Mills coupling $\lambda R\gg1$ while dimensionless charge $q$
is kept finite, or equivalently both chemical potential $\mu$ and
charge density $\rho$ are very small compared to temperature $T$.
In this limit, the back-reactions from the gauge fields to the metric
is negligible at the leading order, so the background metric is an
AdS-Schwarzschild black hole. But here, we will propose a more refined
perturbative approach based on variational method to solve the system
analytically for finite $\lambda R$ near the critical regime. The
advantages of our approach are that it is a systematic method to go
beyond the Probe limit and compute all the back-reactions between
metric and gauge fields, and it is extremely suitable to be directly
applied on the analytic computation of Hall viscosity and its ratio
to angular momentum density, as will be shown in the next two sections. 

Before proceeding to the actual calculation, we will first make a
note here about how we will present our finite $\lambda R$ results
in this paper. In our calculations we don't make any assumptions about
how big or small $\lambda R$ is. But the expressions we obtain from
the variational method for finite $\lambda R$ are usually very complicated
and not illuminating. To present them in a better way, for every result
we will make a large $\lambda R$ series expansion and keep only the
first two leading terms. The leading terms are the same as one can
get from the Probe limit; the next-to-leading terms can only be obtained
by fully taking into account the back-reactions. The next-to-next-to-leading
order terms are usually of order $O\left(\frac{1}{(\lambda R)^{4}}\right)$
compared to the leading terms. This means that even for, say $\lambda R=2$,
the relative error due to the series expansion is just about $6\%$.
So it is reasonable to believe that results presented in this way
are not only accurate for $\lambda R\gg1$, but also very good down
to $\lambda R\sim O\left(1\right)$. The regime where $\lambda R$
is really small and the above series expansion can not hold corresponds
to the low temperature limit, and we will deal this regime separately
in a late section at the end of this paper. Thus we will eventually
get a qualitatively complete result for all possible range of $\lambda R$.

\subsection{Critical Line and Phase Diagram}

The first step in our analytic approach is to find the expression
for the critical temperature $T_{c}$ in terms of other physical parameters,
i.e. the phase diagram. 

To start, notice that at $T=T_{c}$ the symmetry-breaking solution
with $A(z)\neq0$ transits continuously to the AdS-Reissner-Nordström
solution with $A(z)=0$. So when $T$ is below but very close to $T_{c}$,
$A(z)$ is very small and can be treated perturbatively, thus the
non-linear equations (\ref{EQ1})-(\ref{EQ5}) can be linearized and
solved order by order. Let $\epsilon\ll1$ be a book-keeping parameter
of the perturbative expansion. It marks the ``smallness'' of $A(z)$
near the critical temperature, thus marks the deviation from the AdS-Reissner-Nordström
solution (\ref{AdSRN Solution}) order by order, and at the end we
will always set $\epsilon=1$. The ansatz for the background fields
is: 
\begin{equation}
\left\{ \begin{aligned} & A(z)=A^{(1)}(z)\epsilon+A^{(2)}(z)\epsilon^{3}+O\left(\epsilon^{5}\right)\\
 & r(z)=r^{(0)}(z)+r^{(1)}(z)\epsilon^{2}+r^{(2)}(z)\epsilon^{4}+O\left(\epsilon^{6}\right)\\
 & F(z)=F^{(0)}(z)+F^{(1)}(z)\epsilon^{2}+F^{(2)}(z)\epsilon^{4}+O\left(\epsilon^{6}\right)\\
 & \Phi(z)=\Phi^{(0)}(z)+\Phi^{(1)}(z)\epsilon^{2}+\Phi^{(2)}(z)\epsilon^{4}+O\left(\epsilon^{6}\right)
\end{aligned}
\:.\right.\label{A_Expansion}
\end{equation}
 The first equation to solve is the linearized equation for $A^{(1)}(z)$
from (\ref{EQ3}): 
\begin{equation}
\frac{d}{dz}\left[F^{(0)}(z)\left(\frac{d}{dz}A^{(1)}(z)\right)\right]+\frac{\lambda^{2}\Phi^{(0)}(z)^{2}}{F^{(0)}(z)}A^{(1)}(z)=0\:.\label{EQ3_1}
\end{equation}
Near horizon $z=z_{H}$, this equation has two characteristic solutions
for $A^{(1)}(z)$: one is regular and the other contains $\ln(z-z_{H})$.
We require $A^{(1)}(z)$ to be regular near horizon, thus choose one
of the two integration constants to kill the $\ln(z-z_{H})$ solution.
Now for generic parameters this completely fixes the solution (particularly
its near-boundary behavior at $z\rightarrow\infty$) up to an overall
normalization constant. This means near the boundary $z\rightarrow\infty$
we have 
\begin{equation}
A^{(1)}(z)=\alpha_{0}^{(1)}+\frac{\alpha_{1}^{(1)}}{z}+O\left(\frac{1}{z^{2}}\right)\:,
\end{equation}
where $\alpha_{0}^{(1)}$ being a fixed function of all parameters
is generally non-vanishing. Thus generally the sourceless condition
(\ref{NoSource}) can not be achieved only at linear level; its full
realization requires the inclusion of higher order terms through the
nonlinear term in (\ref{EQ3}), and this will also determine the behavior
of the order parameter $\alpha_{1}^{(1)}$. However, there are special
cases that the sourceless condition (\ref{NoSource}) can be achieved
at linear level, when the parameters take some special discrete values.
This corresponds to an eigenvalue problem for (\ref{EQ3_1}), and
the eigensolution with the highest eigenvalue of $T$ (thus the lowest
eigenvalue of $q$) corresponds precisely to the case $T=T_{c}$.
So we have the condition for the critical temperature: 
\begin{equation}
\alpha_{0}^{(1)}(T=T_{c},\lambda,\ldots)=0\:.\label{Tc Condition}
\end{equation}
The above equation gives the expression for $T_{c}$ in terms of the
other physical parameters $\lambda,\ldots$, thus the phase diagram.

Given the complicated form of (\ref{AdSRN Solution}), (\ref{EQ3_1})
can not be solved analytically in terms of special functions. We use
variational method to solve it. Even though this is an approximate
method, we will later see that its accuracy is surprisingly high.
Solving (\ref{EQ3_1}) with appropriate boundary conditions corresponds
to finding the extrema of the action 
\begin{equation}
I^{(1)}=\int_{z_{H}}^{\infty}dz\left[-F^{(0)}(z)\left(\frac{d}{dz}A^{(1)}(z)\right)^{2}+\frac{\lambda^{2}\Phi^{(0)}(z)^{2}}{F^{(0)}(z)}A^{(1)}(z)^{2}\right]+I_{\textrm{boundary}}^{(1)}\:,
\end{equation}
where $I_{\textrm{boundary}}^{(1)}$ contains boundary terms such
that under given boundary conditions the variational problem is well-defined
(i.e. the variations of all boundary terms are vanishing). We have
already imposed one regularity condition near the horizon. Given that
the blackening function $F^{(0)}(z)$ is vanishing at the horizon,
this condition does not introduce any boundary term to $I_{\textrm{boundary}}^{(1)}$.
Now we fix the normalization of $A^{(1)}(z)$ by requiring that%
\footnote{Notice that $\alpha_{1}$ is related to the order parameter $\langle\mathcal{O}\rangle$.%
} 
\begin{equation}
\alpha_{1}^{(1)}\textrm{ is fixed.}
\end{equation}
This introduces a boundary term 
\begin{equation}
I_{\textrm{boundary}}^{(1)}=-\frac{2\alpha_{0}^{(1)}\alpha_{1}^{(1)}}{R^{2}}\:.
\end{equation}
 Now we choose a form of the trial function:%
\footnote{The function in the first $\left(\,\right)$ is to cancel the same
factor in the blackening function $F^{(0)}(z)$ in the denominator
of the action such that the integral is easy to do. One can of course
choose other ansatz and will get similar results. $n$ is the rank
of the trial polynomial. In practice one can only include the first
few terms (like to set $n=4\;\textrm{or}\;6$) and will get very accurate
results. We find that choosing $n$ to be an even number usually gives
better results. In this paper, all calculations are done by setting
$n=4$.%
} 
\begin{equation}
A^{(1)}(z)=\alpha_{1}^{(1)}\left(1+\frac{z_{H}}{z}+\frac{z_{H}^{2}}{z^{2}}-\frac{q^{2}z_{H}^{3}}{4\lambda^{2}R^{2}z^{3}}\right)\left(c_{0}^{(1)}+\frac{1-z_{H}c_{0}^{(1)}}{z}+\sum_{i=2}^{n}\frac{c_{i}^{(1)}}{z^{2}}\right)\:,\label{A1_trial}
\end{equation}
and then compute the action $I^{(1)}$ and solve $c_{i}^{(1)}$ by
minimizing it: 
\begin{equation}
\frac{\partial I^{(1)}}{\partial c_{i}^{(1)}}=0\quad(i=0,2,3,\ldots,n)\:.
\end{equation}
Notice that given the above trial ansatz, 
\begin{equation}
\alpha_{0}^{(1)}=c_{0}^{(1)}\alpha_{1}^{(1)}
\end{equation}
and the condition for critical temperature (\ref{Tc Condition}) becomes
$c_{0}^{(1)}(T=T_{c},\lambda,\ldots)=0$. We will not give the full
expressions for $c_{i}^{(1)}$, since they are messy and interested
readers can easily repeat the calculation. The numerator of $c_{0}^{(1)}$
is a bi-polynomial of $q$ and $\lambda R$ with high ranks, and finding
the critical temperature according to the condition (\ref{Tc Condition})
corresponds to finding the roots of this polynomial: $q=q_{\textrm{root}}(\lambda R)$.
However, even for the simplest ansatz the polynomial usually goes
beyond rank-5 thus it does not have an explicit analytic expression
for its roots. But numeric plot shows that these roots all have similar
and simple behaviors in $(q,\lambda R)$-plane: they start at the
origin, go closely along the extremal lines (\ref{Extremal Limit})
and then at some points turn rapidly to constant-$q$ lines. Thus
before the turning point, the solutions are essentially $T=0$. After
the turning point, to obtain the constant-$q$ solutions, we can take
the $\lambda R\gg1$ limit in the numerator of $c_{0}^{(1)}$ and
keeping only the first two leading terms will give good enough results.
The actual critical temperature corresponds to the smallest $q_{\textrm{root}}$:
\begin{equation}
q_{c}=3.69-\frac{0.662}{(\lambda R)^{2}}+O\left(\frac{1}{(\lambda R)^{4}}\right)\:.\label{Critical Line}
\end{equation}
To obtain the expression for critical temperature $T_{c}$, we need
to use the ensemble conditions (\ref{EnsembleCondition}) to convert
$z_{H}$ in (\ref{Temperature}) to physical parameters. For Canonical
ensemble we have 
\begin{equation}
\frac{T_{c}}{\sqrt{\hat{\rho}}}=1.96\sqrt{\lambda R}\left[1-\frac{1.04}{(\lambda R)^{2}}+O\left(\frac{1}{(\lambda R)^{4}}\right)\right]\theta\left(\lambda-\lambda_{c}\right)\:,\label{Tc_Canonical}
\end{equation}
where $\hat{\rho}=\frac{\kappa^{2}}{(2\pi)^{3}R^{2}}\rho$ as defined
in \cite{Gubser:2008zu}. For Grand Canonical ensemble we have 
\begin{equation}
\frac{T_{c}}{\mu}=0.129\lambda R\left[1-\frac{0.954}{(\lambda R)^{2}}+O\left(\frac{1}{(\lambda R)^{4}}\right)\right]\theta\left(\lambda-\lambda_{c}\right)\:,\label{Tc_Grand}
\end{equation}
where $\theta(x)$ is the Heaviside step function. From the above
two expressions we can see that $T_{c}$ will reach zero at a critical
coupling around $\lambda_{c}R\approx1$. We will deal with $T\rightarrow0$
limit separately later to give a more accurate expression for this
critical coupling $\lambda_{c}$. The phase diagrams are shown in
Figure \ref{Fig:PhaseDiagrams}. Comparing with the numeric plot FIG.1(A)
and equation (B8) in \cite{Gubser:2008zu} we see quantitatively they
are almost the same. Away from but close to the critical line (\ref{Critical Line})
we find 
\begin{eqnarray}
c_{0}^{(1)} & = & \frac{1}{\kappa R^{\frac{3}{2}}\sqrt{\rho\lambda}}\left[-2.10+\frac{8.31}{(\lambda R)^{2}}+O\left(\frac{1}{(\lambda R)^{4}}\right)\right]\left(1-\frac{T}{T_{c}}\right)\theta\left(T_{c}-T\right)\nonumber \\
 & = & \frac{1}{\mu\lambda R^{3}}\left[-1.01+\frac{1.80}{(\lambda R)^{2}}+O\left(\frac{1}{(\lambda R)^{4}}\right)\right]\left(1-\frac{T}{T_{c}}\right)\theta\left(T_{c}-T\right)\:,\label{c0_1}
\end{eqnarray}
where the first line is for Canonical ensemble and second line Grand
Canonical ensemble. This result will be very useful later.

\begin{center}
\begin{figure}[t]
\begin{centering}
\includegraphics[width=7cm]{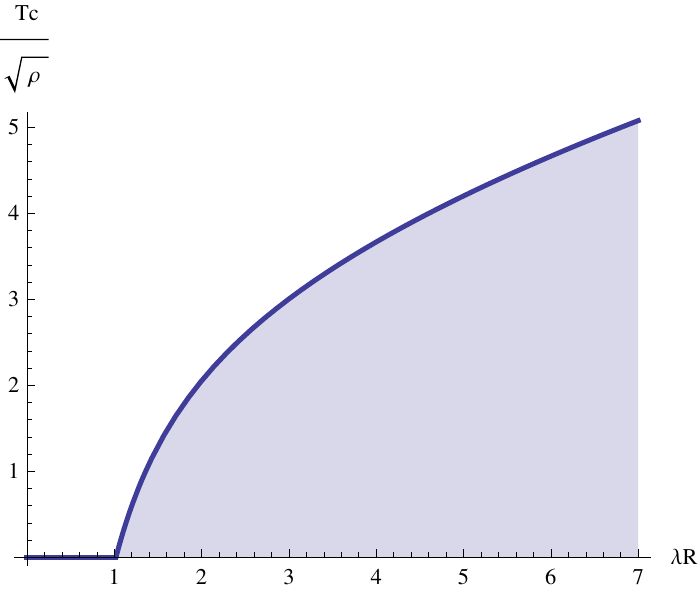}\qquad{} \includegraphics[width=7cm]{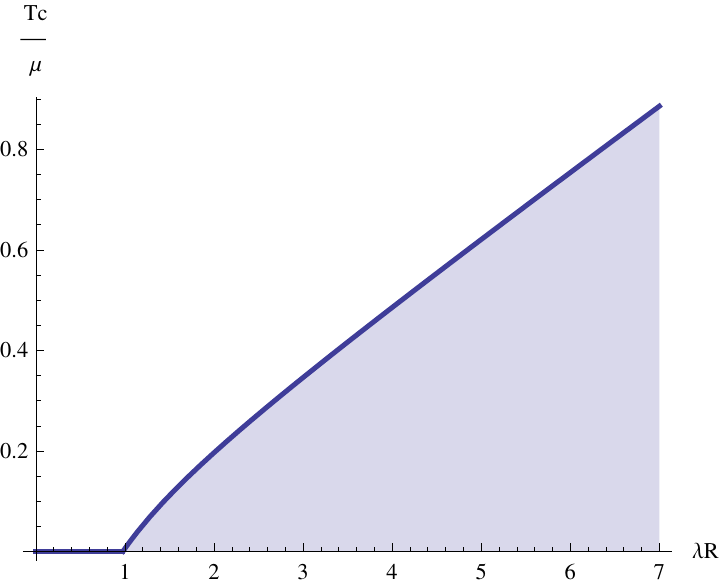}
\par\end{centering}

\caption{Phase diagrams for Canonical ensemble (left) and Grand Canonical ensemble
(right). The blue lines are the critical lines given by equations
(\ref{Tc_Canonical}) and (\ref{Tc_Grand}). The shaded areas are
the superconducting/superfluid phase and the unshaded areas the normal
phase.}
\label{Fig:PhaseDiagrams}
\end{figure}

\par\end{center}

\subsection{Near-Critical Behavior of Order Parameter}

Now going back to the boundary condition (\ref{Tc Condition}), we
can see away from but close to $T_{c}$, we have 
\begin{equation}
\alpha_{0}^{(1)}{\Big|_{T\rightarrow T_{c}}}=-a^{(1)}\alpha_{1}^{(1)}\left(T_{c}-T\right)\:,\label{alpha1_0}
\end{equation}
where $a^{(1)}$ is a positive quantity one can retrieve from (\ref{c0_1}).
The fulfillment of the sourceless condition (\ref{NoSource}) now
requires to include the next order solution: 
\begin{equation}
\alpha_{0}^{(1)}+\alpha_{0}^{(2)}{\Big|_{T\rightarrow T_{c}}}=0\:,\label{NoSource_2}
\end{equation}
where near the boundary 
\begin{equation}
A^{(2)}(z)=\alpha_{0}^{(2)}+\frac{\alpha_{1}^{(2)}}{z}+O\left(\frac{1}{z^{2}}\right)\:.\label{A2_NB}
\end{equation}

To proceed, we need to solve the other first order fields $r^{(1)}(z)$,
$F^{(1)}(z)$ and $\Phi^{(1)}(z)$ first. They satisfy a set of coupled
second order inhomogeneous linear equations derived from (\ref{EQ1})-(\ref{EQ4}),
with all source terms quadratic in $A^{(1)}(z)$. These equations
can be all put to integrable forms using the $O\left(\epsilon^{2}\right)$
order of the trace equation (\ref{EQ5}) and background equations
which (\ref{AdSRN Solution}) satisfies, and then be integrated out
one by one. The results are listed in Appendix (C), with six arbitrary
integration constants $C_{1}$-$C_{6}$ to be fixed by appropriate
boundary conditions. $C_{1}$ is fixed by requiring that $r(z)$ goes
asymptotically AdS near the boundary, that is, there is no $z$ term
in $r^{(1)}(z)$ when $z\rightarrow\infty$. We also require $F(z)\rightarrow4\pi T(z-z_{H})$
near the horizon, which means both the horizon position $z_{H}$ and
the AdS-RN temperature-charge relation (\ref{Temperature}) are unchanged
in the presence of the condensate $A^{(1)}(z)$. This implies $F^{(1)}(z)\rightarrow O\left((z-z_{H})^{2}\right)$,
which fixes $C_{5}$ and $C_{6}$ in terms of the other constants
through the vanishing of constant and linear terms near horizon. Notice
that the asymptotic AdS requirement for $F^{(1)}(z)$ is automatically
satisfied, thus fixes no more constant. The $O\left(\epsilon^{2}\right)$
order of the trace equation listed in Appendix (C) fixes the constant
$C_{2}$ in terms of the remaining. When solving this equation, one
shall bear in mind that the equation (\ref{EQ3_1}) is not solved
exactly, but just approximately by variational method. The consequence
of that is that the $O\left(\epsilon^{2}\right)$ order trace equation
will never be solved exactly either. To avoid this complication, we
shall not require that the whole trace equation hold; instead, we
will just look at the near boundary leading order of its left hand
side and require this term alone to vanish. It is not hard to see
that all the non-vanishing near boundary sub-leading orders we have
omitted here are a consequence of (\ref{EQ3_1}) being solved approximately,
and if (\ref{EQ3_1}) was solved exactly, they will all vanish automatically.
Now we are left with two constants $C_{3}$ and $C_{4}$ to be fixed
by conditions for $\Phi^{(1)}(z)$. One condition is obvious: we require
$\Phi^{(1)}(z)\rightarrow O\left(z-z_{H}\right)$ near horizon such
that $\Phi(z)\rightarrow O\left(z-z_{H}\right)$ in (\ref{NH Conditions})
holds. The vanishing of the near-horizon constant term fixes one of
$C_{3}$ and $C_{4}$ in term of the other. By now, we have essentially
exhausted all the boundary conditions we \emph{must} impose for $r^{(1)}(z)$,
$F^{(1)}(z)$ and $\Phi^{(1)}(z)$ that are consequences of (\ref{NB Conditions})
and (\ref{NH Conditions}), but we are still left with an undetermined
constant, basically a combination of $C_{3}$ and $C_{4}$. This is
not surprising, because this is just a reflection of gauge freedoms
related to $\Phi^{(1)}(z)$ and other fields at this order: the undetermined
constant is associated with a pure-gauge solution, and this constant,
if kept arbitrary, will not appear in any physical results that are
expressed in terms of \emph{physical} variables such as $T$, $\rho$,
$\mu$ and $\lambda$ (note that $q$ and $z_{H}$ are \emph{not}
physical variables). We have explicitly verified this claim by keeping
this constant arbitrary in all follow-up computations. But to make
the computations more compact and transparent, we can use the gauge
freedom to fix this arbitrary constant. There are two natural choices:
if we choose to work in Canonical ensemble, in which $\rho\sim\Phi_{1}$
is kept fixed, we will hope that $\Phi_{1}$ is not altered in the
presence of $A^{(1)}(z)$, thus we can require $\Phi^{(1)}(z)\rightarrow\textrm{const.}+O\left(z^{-2}\right)$,
i.e. $\Phi_{1}=\Phi_{1}^{(0)}$ and $\Phi_{1}^{(1)}=0$; on the other
hand, if we choose to work in Grand Canonical ensemble, in which $\mu\sim\Phi_{0}$
is fixed, we can require $\Phi^{(1)}(z)\rightarrow O\left(z^{-1}\right)$,
i.e. $\Phi_{0}=\Phi_{0}^{(0)}$ and $\Phi_{0}^{(1)}=0$, which means
$\Phi_{0}$ is not altered in the presence of $A^{(1)}(z)$. Both
choices, among others, shall give the same physical results at the
end.

Since the actual calculation based on $A^{(1)}(z)$ obtained from
variational method gives extremely baroque expressions for $r^{(1)}(z)$,
$\Phi^{(1)}(z)$ and $F^{(1)}(z)$ (as well as all second order fields
to be discussed later), we will not give their explicit expressions
in this paper. Interested readers should be able to repeat our calculation
easily following the procedures we have outlined here. We will only
list some simple key results derived from them.

We are now at a position to solve second order fields. $A^{(2)}(z)$
satisfies 
\begin{equation}
\frac{d}{dz}\left[F^{(0)}(z)\left(\frac{d}{dz}A^{(2)}(z)\right)\right]+\frac{\lambda^{2}\Phi^{(0)}(z)^{2}}{F^{(0)}(z)}A^{(2)}(z)+\Upsilon^{(2)}(z)=0\:,
\end{equation}
 where 
\begin{eqnarray}
\Upsilon^{(2)}(z) & = & \left(\frac{d}{dz}A^{(1)}(z)\right)\left[\left(\frac{d}{dz}F^{(1)}(z)\right)-\frac{F^{(1)}(z)}{F^{(0)}(z)}\left(\frac{d}{dz}F^{(0)}(z)\right)\right]\nonumber \\
 &  & -\lambda^{2}A^{(1)}(z)\left(\frac{A^{(1)}(z)^{2}}{r^{(0)}(z)^{2}}+2\frac{\Phi^{(0)}(z)^{2}F^{(1)}(z)}{F^{(0)}(z)^{2}}-2\frac{\Phi^{(0)}(z)\Phi^{(1)}(z)}{F^{(0)}(z)}\right)
\end{eqnarray}
 is a known source function in terms of the above solved first order
fields. The action associated with the variational problem is 
\begin{equation}
I^{(2)}=\int_{z_{H}}^{\infty}dz\left[-F^{(0)}(z)\left(\frac{d}{dz}A^{(2)}(z)\right)^{2}+\frac{\lambda^{2}\Phi^{(0)}(z)^{2}}{F^{(0)}(z)}A^{(2)}(z)^{2}+2\Upsilon^{(2)}(z)A^{(2)}(z)\right]+I_{\textrm{boundary}}^{(2)}\:.
\end{equation}
Since $A^{(2)}(z)$ satisfies an inhomogeneous equation with a source
$\Upsilon^{(2)}(z)$, we can impose two boundary conditions to completely
determine $A^{(2)}(z)$ in term of $\Upsilon^{(2)}(z)$: one is again
the regularity condition near the horizon, and the second is chosen
to be 
\begin{equation}
\alpha_{1}^{(2)}=0\:,
\end{equation}
where $\alpha_{1}^{(2)}$ is defined in (\ref{A2_NB}). Actually in
the perturbative expansion of $A(z)$, the splitting between $A^{(1)}(z)$
and $A^{(2)}(z)$ is arbitrary: one can always take a small part of
$A^{(1)}(z)$ which is of the same order of $A^{(2)}(z)$ and sneak
it into the latter and the perturbative expansion still holds. The
meaning of the above boundary condition is just to make the splitting
unique, or equivalently one can think it is the definition of $A^{(1)}(z)$.
The above boundary condition introduces no more boundary term since
itself is vanishing, thus 
\begin{equation}
I_{\textrm{boundary}}^{(2)}=0\:.
\end{equation}
 Using a trial ansatz 
\begin{equation}
A^{(2)}(z)=\left(1+\frac{z_{H}}{z}+\frac{z_{H}^{2}}{z^{2}}-\frac{q^{2}z_{H}^{3}}{4\lambda^{2}R^{2}z^{3}}\right)\left(\alpha_{0}^{(2)}-\frac{z_{H}\alpha_{0}^{(2)}}{z}+\sum_{i=2}^{n}\frac{c_{i}^{(2)}}{z^{2}}\right)
\end{equation}
 we can solve $\alpha_{0}^{(2)}$ and $c_{i}^{(2)}$ in a similar
fashion as before. Since the source $\Upsilon^{(2)}(z)$ is cubic
in $A^{(1)}(z)$ thus cubic in $\alpha_{1}^{(1)}$, it's not hard
to see that close to $T_{c}$ 
\begin{equation}
\alpha_{0}^{(2)}{\Big|_{T\rightarrow T_{c}}}=a^{(2)}\left(\alpha_{1}^{(1)}\right)^{3}\:.
\end{equation}
$a^{(2)}$ is a positive quantity in terms of other parameters: 
\begin{eqnarray}
a^{(2)} & = & \frac{1}{\kappa^{5}R^{\frac{7}{2}}\rho^{\frac{5}{2}}\sqrt{\lambda}}\left[1.44-\frac{3.98}{(\lambda R)^{2}}+O\left(\frac{1}{(\lambda R)^{4}}\right)\right]\nonumber \\
 & = & \frac{1}{R^{11}\mu^{5}\lambda^{3}}\left[0.688-\frac{0.783}{(\lambda R)^{2}}+O\left(\frac{1}{(\lambda R)^{4}}\right)\right]\:,\label{a_2}
\end{eqnarray}
where the first line is for Canonical ensemble and the second line
Grand Canonical ensemble. Now the sourceless condition (\ref{NoSource_2})
reads 
\begin{equation}
-a^{(1)}\alpha_{1}^{(1)}\left(T_{c}-T\right)+a^{(2)}\left(\alpha_{1}^{(1)}\right)^{3}=0\:.
\end{equation}
Since both $a^{(1)}$ and $a^{(2)}$ are positive,%
\footnote{This assertion for $a^{(2)}$ need to be made carefully, and we will
have more discussion on this later for low temperature limit.%
} when $T>T_{c}$ there is only one trivial solution $\alpha_{1}^{(1)}=0$,
which means $A(z)=0$. This reproduces the AdS-RN solution and shows
it is the only possible solution when $T>T_{c}$. When $T<T_{c}$
there is another non-trivial solution 
\begin{equation}
\alpha_{1}^{(1)}=\left(\frac{a^{(1)}}{a^{(2)}}\right)^{\frac{1}{2}}\sqrt{T_{c}-T}\qquad\quad(T<T_{c})\:,
\end{equation}
which yields a non-vanishing $A(z)$. Notice that $\alpha_{1}$ is
proportional to the order parameter $\langle\mathcal{O}\rangle$,
we obtain $\langle\mathcal{O}\rangle\sim\sqrt{T_{c}-T}\theta\left(T_{c}-T\right)$,
which agrees with Ginzburg-Landau theory. For Canonical ensemble,
we get%
\footnote{Compare to the numerical result in \cite{Roberts:2008ns}: converting
their equation (16) to our conventions,and noticing their definition
for $\langle J\rangle$ involves a factor of $\sqrt{2}$, they have
\[
\frac{\langle\mathcal{O}\rangle}{\rho}=1.16\left(1-\frac{T}{T_{c}}\right)^{\frac{1}{2}}\theta\left(T_{c}-T\right)
\]
at the leading order. So the agreement is good.%
} 
\begin{equation}
\frac{\langle\mathcal{O}\rangle}{\rho}=1.21\left[1-\frac{0.594}{(\lambda R)^{2}}+O\left(\frac{1}{(\lambda R)^{4}}\right)\right]\left(1-\frac{T}{T_{c}}\right)^{\frac{1}{2}}\theta\left(T_{c}-T\right)\:.\label{eq:Condensate_Canon}
\end{equation}
For Grand Canonical ensemble 
\begin{equation}
\frac{\langle\mathcal{\hat{O}}\rangle}{\mu^{2}}=4.88\times10^{-3}\lambda R\left[1-\frac{0.322}{(\lambda R)^{2}}+O\left(\frac{1}{(\lambda R)^{4}}\right)\right]\left(1-\frac{T}{T_{c}}\right)^{\frac{1}{2}}\theta\left(T_{c}-T\right)\:,\label{eq:Condensate_GrandCanon}
\end{equation}
 where $\langle\mathcal{\hat{O}}\rangle=\frac{\kappa^{2}}{(2\pi)^{3}R^{2}}\langle\mathcal{O}\rangle$
as defined in \cite{Gubser:2008zu}.

\subsection{Near-Critical Behaviors of Charge Density and Chemical Potential}

We can also compute the change of chemical potential $\mu$ or charge
density $\rho$, when the other one is fixed, as a function of $T_{c}-T$
near the critical temperature. There are two contributions to it.
The first one is from the AdS-RN part $\Phi^{(0)}$, which we can
obtain by directly vary (\ref{AdSRN Thermal Relation}), and it is
proportional to $T_{c}-T$. The second contribution is from $\Phi^{(1)}$,
by computing either $\Phi_{1}^{(1)}$ for $\rho$ or $\Phi_{0}^{(1)}$
for $\mu$, which is proportional to $\langle\mathcal{O}\rangle^{2}$,
thus also to $T_{c}-T$. So we can see that they are of the same order.
For Canonical ensemble, we get 
\begin{equation}
\mu-\mu_{c}=0.165\sqrt{\frac{\rho}{\lambda}}\frac{\kappa}{R^{\frac{3}{2}}}\left[1-\frac{1.63}{(\lambda R)^{2}}+O\left(\frac{1}{(\lambda R)^{4}}\right)\right]\left(1-\frac{T}{T_{c}}\right)\theta\left(T_{c}-T\right)\:,
\end{equation}
where 
\begin{equation}
\mu_{c}(\rho,\lambda)=0.960\sqrt{\frac{\rho}{\lambda}}\frac{\kappa}{R^{\frac{3}{2}}}\left[1-\frac{0.0898}{(\lambda R)^{2}}+O\left(\frac{1}{(\lambda R)^{4}}\right)\right]\:.
\end{equation}
For Grand Canonical ensemble we get 
\begin{equation}
\rho-\rho_{c}=-0.318\lambda\mu^{2}\frac{R^{3}}{\kappa^{2}}\left[1-\frac{1.04}{(\lambda R)^{2}}+O\left(\frac{1}{(\lambda R)^{4}}\right)\right]\left(1-\frac{T}{T_{c}}\right)\theta\left(T_{c}-T\right)\:,
\end{equation}
where $\rho_{c}(\mu,\lambda)$ can be obtained by inverting the above
expression of $\mu_{c}(\rho,\lambda)$.

Now we can rewrite $\langle\mathcal{O}\rangle$ in terms of $\mu-\mu_{c}$
and $\rho-\rho_{c}$: 
\begin{eqnarray}
\langle\mathcal{O}\rangle & = & 2.98\lambda^{\frac{1}{4}}\rho^{\frac{3}{4}}\frac{R^{\frac{3}{4}}}{\sqrt{\kappa}}\left[1+\frac{0.220}{(\lambda R)^{2}}+O\left(\frac{1}{(\lambda R)^{4}}\right)\right]\sqrt{\mu-\mu_{c}}\theta\left(\mu-\mu_{c}\right)\nonumber \\
 & = & 2.15\sqrt{\lambda}\mu\frac{R^{\frac{3}{2}}}{\kappa}\left[1+\frac{0.197}{(\lambda R)^{2}}+O\left(\frac{1}{(\lambda R)^{4}}\right)\right]\sqrt{\rho_{c}-\rho}\theta\left(\rho_{c}-\rho\right)\:,
\end{eqnarray}
where the first line is for Canonical ensemble and second line Grand
Canonical ensemble. The critical line has 
\begin{equation}
\mu-\mu_{c}=\frac{\partial\mu_{c}}{\partial\rho}\left(\rho-\rho_{c}\right)=\frac{0.480}{\sqrt{\lambda\rho}}\left[1-\frac{0.0898}{(\lambda R)^{2}}+O\left(\frac{1}{(\lambda R)^{4}}\right)\right]\frac{\kappa}{R^{\frac{3}{2}}}\left(\rho-\rho_{c}\right)\:.\label{CriticalDeriv}
\end{equation}
It is straightforward to check that if the calculation is consistent,
then the ratio of the first line over the second line in the above
expression of $\langle\mathcal{O}\rangle$ shall be $1$ if we substitute
in the absolute value of (\ref{CriticalDeriv}). We get 
\begin{equation}
1.00+\frac{0.067}{(\lambda R)^{2}}+O\left(\frac{1}{(\lambda R)^{4}}\right)\:.
\end{equation}
We believe the $\lambda R$ term is a numeric error because the variational
method is approximate. So within this error our calculation for the
two ensembles is consistent.

\subsection{Near-Critical Behaviors of Characteristic Functions}

Next it is straightforward to solve other second order fields $r^{(2)}(z)$,
$F^{(2)}(z)$ and $\Phi^{(2)}(z)$ following the same procedures for
first order fields as discussed before, then use (\ref{FreeEnergyExpression})
and (\ref{GrandPotentialExpression}) to compute characteristic functions
for the two ensembles. For Canonical ensemble, we find the free energy
density of the unbroken phase (pure AdS-RN background) is 
\begin{equation}
f_{\textrm{Helmholtz}}^{(0)}=-\frac{32\pi^{3}R^{2}}{27\kappa^{2}}T^{3}+\frac{3\kappa^{2}}{8\pi R^{2}}\frac{\rho^{2}}{T}+O\left(\frac{1}{\lambda^{2}}\right)
\end{equation}
and the free energy density difference between the broken phase ($A(z)\neq0$)
and unbroken phase near the critical temperature is 
\begin{equation}
\Delta f_{\textrm{Helmholtz}}=-\frac{6.45}{\lambda R}\rho T_{c}\left(1-\frac{T}{T_{c}}\right)^{2}\theta\left(T_{c}-T\right)+O\left(\frac{1}{\lambda^{3}}\right)\:,
\end{equation}
while for Grand Canonical ensemble the grand potential density of
the unbroken phase is 
\begin{equation}
\Omega_{\textrm{Grand}}^{(0)}=-\frac{32\pi^{3}R^{2}}{27\kappa^{2}}T^{3}+O\left(\frac{1}{\lambda^{2}}\right)
\end{equation}
and the grand potential density difference between the broken and
unbroken phase near the critical temperature is 
\begin{equation}
\Delta\hat{\Omega}_{\textrm{Grand}}=-1.61\times10^{-3}\lambda R\mu^{3}\left(1-\frac{T}{T_{c}}\right)^{2}\theta\left(T_{c}-T\right)+O\left(\frac{1}{\lambda}\right)\:,
\end{equation}
where $\hat{\Omega}_{\textrm{Grand}}=\frac{\kappa^{2}}{(2\pi)^{3}R^{2}}\Omega_{\textrm{Grand}}$.
Notice that in both cases, below $T_{c}$ the broken phase has lower
free energy (grand potential) density than the unbroken phase, so
it is the preferred phase and the phase transition can indeed happen;
and the fact that the characteristic functions are quadratic in $T_{c}-T$
indicates that this is a second order phase transition.

\bigskip{}

\section{Tensor Mode Fluctuations and Viscosities}

\subsection{Mode Classification}

We now look at metric and gauge field fluctuations in the background
(\ref{BackgroundAnsatz}). Their EOMs are given by (\ref{LinearEinstein})
and (\ref{LinearYangMills}). We will work in momentum space where
coordinates $(t,x,y)$ are Fourier transformed to momentum $(\omega,k_{x},k_{y})$.
All fluctuations in (\ref{LinearEinstein}) and (\ref{LinearYangMills})
are highly coupled with each others, mainly because the joint $U(1)$
symmetry in $(x,y)$-plane and $(\mathbf{1},\mathbf{2})$-vector space
is now broken explicitly by any non-zero spatial momentum vector $\vec{k}=(k_{x},k_{y})\neq0$.
In this case the usual classification of tensor, vector (shear) and
scalar (sound) modes are not very helpful because they do not decouple
from each other. However, since we are only interested in viscosities,
which are only related to $\omega$ terms in the correlation functions,
we can assume the fluctuations to be spatially homogeneous, i.e. $\vec{k}=0$.
Now the joint $U(1)$ symmetry is respected and we find there are
indeed three decoupled modes, which behave as tensor, vector and scalar
under the joint rotation of both $(x,y)$-plane and $(\mathbf{1},\mathbf{2})$-vector
space by the same angle:
\begin{itemize}
\item Tensor mode: \{$h_{xy}$, $a_{y}^{\mathbf{1}}+a_{x}^{\mathbf{2}}$\},
{[}$h_{xx}-h_{yy}$, $a_{x}^{\mathbf{1}}-a_{y}^{\mathbf{2}}${]};
\item Vector mode: \{$h_{tx}+h_{ty}$, $h_{xz}+h_{yz}$, $a_{t}^{\mathbf{1}}+a_{t}^{\mathbf{2}}$,
$a_{z}^{\mathbf{1}}+a_{z}^{\mathbf{2}}$, $a_{x}^{\mathbf{3}}+a_{y}^{\mathbf{3}}$\},
{[}$h_{tx}-h_{ty}$, $h_{xz}-h_{yz}$, $a_{t}^{\mathbf{1}}-a_{t}^{\mathbf{2}}$,
$a_{z}^{\mathbf{1}}-a_{z}^{\mathbf{2}}$, $a_{x}^{\mathbf{3}}-a_{y}^{\mathbf{3}}${]};
\item Scalar mode: $h_{tt}$, $h_{zz}$, $h_{zt}$, $h_{xx}+h_{yy}$, $a_{t}^{\mathbf{3}}$,
$a_{z}^{\mathbf{3}}$, $a_{x}^{\mathbf{1}}+a_{y}^{\mathbf{2}}$, $a_{y}^{\mathbf{1}}-a_{x}^{\mathbf{2}}$.
\end{itemize}
The EOM for each combination of fields listed above is just the same
combination of corresponding components of (\ref{LinearEinstein})
or (\ref{LinearYangMills}). Notice that above in both tensor and
vector mode we group the fields into two brackets: the ones in ``\{
\}'' are all even under the joint parity operation - the simultaneously
exchange of $x\leftrightarrow y$ and $\mathbf{1}\leftrightarrow\mathbf{2}$,
while those in ``{[} {]}'' are all odd. When this joint parity is
a symmetry of the system, the fields in each bracket do not mix with
those in the other bracket, thus inside each of tensor and vector
modes there are two decoupled sub-modes which are eigenstates of the
joint parity. However, here the non-Abelian coupling $\lambda$ explicitly
breaks the joint parity, so fields in two brackets do mix, and one
can expect that the mixing terms must all proportional to $\lambda$.

\subsection{Tensor Mode EOMs}

We now focus on tensor mode fluctuations, through which we will obtain
viscosities. For concreteness, we define shorthand notations 
\begin{align}
h_{xy}\equiv r(z)^{2}h_{e}(\omega,z)\:,\quad\quad & \quad\quad\frac{1}{2}\left(a_{y}^{\mathbf{1}}+a_{x}^{\mathbf{2}}\right)\equiv a_{e}(\omega,z)\:,\\
\frac{1}{2}\left(h_{xx}-h_{yy}\right)\equiv r(z)^{2}h_{o}(\omega,z)\:,\quad\quad & \quad\quad\frac{1}{2}\left(a_{x}^{\mathbf{1}}-a_{y}^{\mathbf{2}}\right)\equiv a_{o}(\omega,z)\:,
\end{align}
where the subscript $e$ or $o$ means whether that field is even
or odd under the joint parity operation. Using background equations
(\ref{EQ1})-(\ref{EQ5}), the equations that the above fields satisfy
can be put into the following forms: 
\begin{eqnarray}
\frac{d}{dz}\left[r(z)^{2}F(z)\left(\frac{d}{dz}h_{i}(\omega,z)\right)\right] & = & S_{i}^{h}(\omega,z;\lambda)\:,\label{EOM_h_tensor}\\
\frac{d}{dz}\left[F(z)\left(\frac{d}{dz}a_{i}(\omega,z)\right)\right]+\frac{\lambda^{2}\Phi(z)^{2}}{F(z)}a_{i}(\omega,z) & = & S_{i}^{a}(\omega,z;\lambda)\:,\label{EOM_a_tensor}
\end{eqnarray}
where $i,\, j=e,\, o$%
\footnote{In this section we will reserve the letters $i$ and $j$ exclusively
for $e$ and $o$.%
} and the sources are 
\begin{eqnarray}
S_{i}^{h}(\omega,z;\lambda) & = & -\left[\omega^{2}\frac{r(z)^{2}}{F(z)}+\frac{\lambda^{2}\Phi(z)^{2}}{F(z)}A(z)^{2}-F(z)\left(\frac{d}{dz}A(z)\right)^{2}\right]h_{i}(\omega,z)\nonumber \\
 &  & +2\left[-F(z)\left(\frac{d}{dz}A(z)\right)\left(\frac{d}{dz}a_{i}(\omega,z)\right)+\frac{\lambda^{2}\Phi(z)^{2}}{F(z)}A(z)a_{i}(\omega,z)\right]\\
 &  & +2i\omega\frac{\lambda\Phi(z)}{F(z)}A(z)\epsilon_{ij}a_{j}(\omega,z)\:,\nonumber \\
S_{i}^{a}(\omega,z;\lambda) & = & \left[F(z)\left(\frac{d}{dz}A(z)\right)\left(\frac{d}{dz}h_{i}(\omega,z)\right)+\frac{\lambda^{2}}{r(z)^{2}}A(z)^{3}h_{i}(\omega,z)\right]\nonumber \\
 &  & -\left(\frac{\omega^{2}}{F(z)}+\frac{\lambda^{2}}{r(z)^{2}}A(z)^{2}\right)a_{i}(\omega,z)\\
 &  & +i\omega\frac{\lambda\Phi(z)}{F(z)}\epsilon_{ij}\left[A(z)h_{j}(\omega,z)-2a_{j}(\omega,z)\right]\:.\nonumber 
\end{eqnarray}
The totally antisymmetric ``tensor'' $\epsilon_{ij}$ is defined
as $\epsilon_{eo}=-\epsilon_{oe}=1$, $\epsilon_{ee}=\epsilon_{oo}=0$
and the repeated subscript $j$ is summed over $e$ and $o$. Since
we are only interested in the hydrodynamic regime where momentum is
small compared to other scales ($\omega\ll2\pi T$), we can treat
$\omega$ perturbatively. Furthermore, we focus on near-critical behaviors
of the transport coefficients, so we can treat $A(z)$ perturbatively
as well. Noticing that every term in $S_{i}^{h}(\omega,z;\lambda)$
and $S_{i}^{a}(\omega,z;\lambda)$ is proportional to either $\omega$
or $A(z)$, all sources can be treated perturbatively. We will expand
all fields $h_{i}(\omega,z)$ and $a_{i}(\omega,z)$ as double series
of $\omega$ and $A(z)$ and (\ref{EOM_h_tensor}) and (\ref{EOM_a_tensor})
can be solved order by order by just integrating the sources. It's
straightforward to integrate (\ref{EOM_h_tensor}), but for (\ref{EOM_a_tensor})
it's not that obvious because the appearance of $a_{i}(\omega,z)$
term on the left hand side spoils the apparent integrability. To integrate
(\ref{EOM_a_tensor}), we need to use its Green's functions (bulk-to-bulk
propagators in the context of holography), which are worked out in
Appendix (A).

\subsection{Boundary-to-Bulk Propagators}

First let us solve the indicial equations of (\ref{EOM_h_tensor})
and (\ref{EOM_a_tensor}) near the boundary and horizon to get the
solution's asymptotic behaviors. Using (\ref{NH Conditions}) in (\ref{EOM_h_tensor})
and (\ref{EOM_a_tensor}) we get 
\begin{equation}
h_{i}(\omega,z),\, a_{i}(\omega,z)\rightarrow(z-z_{H})^{\pm i\frac{\omega}{4\pi T}}\qquad\quad(z\rightarrow z_{H})
\end{equation}
and using (\ref{NB Conditions}) we get 
\begin{equation}
\begin{cases}
h_{i}(\omega,z)\rightarrow1\;\;\textrm{or}\;\; z^{-3}\\
a_{i}(\omega,z)\rightarrow1\;\;\textrm{or}\;\; z^{-1}
\end{cases}\qquad(z\rightarrow\infty)\:.
\end{equation}
We turn on only the $h_{i}$ boundary fields $\bar{h}_{i}$ since
we are only interested in calculating the energy-stress tensor correlators,
so the boundary condition we impose is: 
\begin{equation}
\begin{cases}
h_{i}(\omega,z)\rightarrow\bar{h}_{i}\\
a_{i}(\omega,z)\rightarrow0
\end{cases}\qquad(z\rightarrow\infty)\:.
\end{equation}
Following \cite{Herzog:2002pc,Barnes:2010jp,Arnold:2011ja} we choose
the incoming wave condition near the horizon: 
\begin{equation}
h_{i}(\omega,z),\, a_{i}(\omega,z)\rightarrow(z-z_{H})^{-i\frac{\omega}{4\pi T}}\qquad\quad(z\rightarrow z_{H})\:.
\end{equation}

Following Appendix (A), let us assume that $\Theta_{m}(z)$ ($m=<,>$)
are the two independent solutions to the homogeneous equation associated
with (\ref{EOM_a_tensor}): 
\begin{equation}
\frac{d}{dz}\left[F(z)\left(\frac{d}{dz}\Theta_{m}(z)\right)\right]+\frac{\lambda^{2}\Phi(z)^{2}}{F(z)}\Theta_{m}(z)=0\label{Theta_EOM}
\end{equation}
with the boundary condition 
\begin{equation}
\begin{cases}
\Theta_{<}(z)\textrm{ is regular} & \qquad(z\rightarrow z_{H})\\
\Theta_{>}(z)\rightarrow O\left(z^{-1}\right) & \qquad(z\rightarrow\infty)
\end{cases}\:.
\end{equation}
Notice that given the above boundary condition, usually $\Theta_{<}(z)\rightarrow\textrm{constant}$
when $z\rightarrow\infty$. Assume 
\begin{equation}
\left\{ \begin{aligned} & \Theta_{<}(z)=B_{<}+O\left(\frac{1}{z}\right)\\
 & \Theta_{>}(z)=\frac{B_{>}}{z}+O\left(\frac{1}{z^{2}}\right)
\end{aligned}
\qquad(z\rightarrow\infty)\:,\right.\label{Theta_NB}
\end{equation}
then the normalization constant is 
\begin{equation}
N_{r}=F(z)\textrm{Wr}\left[\Theta_{<}(z),\Theta_{>}(z)\right]=-\frac{B_{<}B_{>}}{R^{2}}\:.\label{Nr_const}
\end{equation}

Now we list the solution to (\ref{EOM_h_tensor}) and (\ref{EOM_a_tensor})
as a double series expansion of small $\omega$ and $A(z)$ up to
orders $O\left(\omega\right)$ and $O\left(A(z)^{4}\right)$: 
\begin{eqnarray}
h_{i}(\omega,z) & = & \left(\frac{z-z_{H}}{z}\right)^{-i\frac{\omega}{4\pi T}}\left\{ \bar{h}_{i}+\sum_{j=1}^{4}h_{i}^{(0j)}(z)+i\omega\left(\sum_{j=0}^{4}h_{i}^{(1j)}(z)\right)+O\left(\omega^{2},A(z)^{5}\right)\right\} \:,\nonumber \\
\label{h_propagator}\\
a_{i}(\omega,z) & = & \left(\frac{z-z_{H}}{z}\right)^{-i\frac{\omega}{4\pi T}}\left\{ \sum_{j=0}^{4}a_{i}^{(0j)}(z)+i\omega\left(\sum_{j=0}^{4}a_{i}^{(1j)}(z)\right)+O\left(\omega^{2},A(z)^{5}\right)\right\} \:,\label{a_propagator}
\end{eqnarray}
where each term carries two superscripts: the first one labels order
in $\omega$ and the second one order in $A(z)$. Many of them are
vanishing and we list below only the non-vanishing ones: 
\begin{eqnarray}
h_{i}^{(02)}(z) & = & \bar{h}_{i}\int_{\infty}^{z}d\xi\frac{A(\xi)}{r(\xi)^{2}}\left(\frac{d}{d\xi}A(\xi)\right)\:,\\
a_{i}^{(03)}(z) & = & \bar{h}_{i}\frac{1}{N_{r}}\Bigg\{-\Theta_{<}(z)\int_{\infty}^{z}d\xi\Theta_{>}(\xi)\frac{A(\xi)}{r(\xi)^{2}}\left[F(\xi)\left(\frac{d}{d\xi}A(\xi)\right)^{2}+\lambda^{2}A(\xi)^{2}\right]\nonumber \\
 &  & \qquad\quad+\Theta_{>}(z)\int_{z_{H}}^{z}d\xi\Theta_{<}(\xi)\frac{A(\xi)}{r(\xi)^{2}}\left[F(\xi)\left(\frac{d}{d\xi}A(\xi)\right)^{2}+\lambda^{2}A(\xi)^{2}\right]\Bigg\}\:,\\
h_{i}^{(04)}(z) & = & \bar{h}_{i}h^{(04)}(z)\:,\\
h_{i}^{(10)}(z) & = & \bar{h}_{i}\left\{ \frac{1}{4\pi T}\ln\left(\frac{z-z_{H}}{z}\right)-r(z_{H})^{2}\int_{\infty}^{z}d\xi\frac{1}{r(\xi)^{2}F(\xi)}\right\} \:,\\
a_{i}^{(11)}(z) & = & \bar{h}_{i}\frac{r(z_{H})^{2}}{N_{r}}\Bigg\{\Theta_{<}(z)\int_{\infty}^{z}d\xi\Theta_{>}(\xi)\frac{1}{r(\xi)^{2}}\left(\frac{d}{d\xi}A(\xi)\right)\nonumber \\
 &  & \qquad\qquad\quad-\Theta_{>}(z)\int_{z_{H}}^{z}d\xi\Theta_{<}(\xi)\frac{1}{r(\xi)^{2}}\left(\frac{d}{d\xi}A(\xi)\right)\Bigg\}\\
 &  & +\epsilon_{ij}\bar{h}_{j}\frac{\lambda}{N_{r}}\left\{ \Theta_{>}(z)\int_{z_{H}}^{z}d\xi\Theta_{<}(\xi)\frac{\Phi(\xi)A(\xi)}{F(\xi)}-\Theta_{<}(z)\int_{\infty}^{z}d\xi\Theta_{>}(\xi)\frac{\Phi(\xi)A(\xi)}{F(\xi)}\right\} \:,\nonumber \\
h_{i}^{(12)}(z) & = & \bar{h}_{i}\Bigg\{\left[\frac{1}{4\pi T}\ln\left(\frac{z-z_{H}}{z}\right)-r(z_{H})^{2}\int_{\infty}^{z}d\rho\frac{1}{r(\rho)^{2}F(\rho)}\right]\int_{\infty}^{z}d\xi\frac{A(\xi)}{r(\xi)^{2}}\left(\frac{d}{d\xi}A(\xi)\right)\nonumber \\
 &  & \qquad\quad+2r(z_{H})^{2}\int_{\infty}^{z}d\rho\frac{1}{r(\rho)^{2}F(\rho)}\int_{z_{H}}^{\rho}d\xi\frac{A(\xi)}{r(\xi)^{2}}\left(\frac{d}{d\xi}A(\xi)\right)\nonumber \\
 &  & \qquad\quad+2\frac{r(z_{H})^{2}}{N_{r}}\Bigg[\int_{\infty}^{z}d\rho\frac{\Theta_{>}(\rho)}{r(\rho)^{2}}\left(\frac{d}{d\rho}A(\rho)\right)\int_{z_{H}}^{\rho}d\xi\frac{\Theta_{<}(\xi)}{r(\xi)^{2}}\left(\frac{d}{d\xi}A(\xi)\right)\nonumber \\
 &  & \qquad\quad-\int_{\infty}^{z}d\rho\frac{\Theta_{<}(\rho)}{r(\rho)^{2}}\left(\frac{d}{d\rho}A(\rho)\right)\int_{\infty}^{\rho}d\xi\frac{\Theta_{>}(\xi)}{r(\xi)^{2}}\left(\frac{d}{d\xi}A(\xi)\right)\Bigg]\Bigg\}\\
 &  & +\epsilon_{ij}\bar{h}_{j}\frac{2\lambda}{N_{r}}\Bigg\{\int_{\infty}^{z}d\rho\frac{\Theta_{<}(\rho)}{r(\rho)^{2}}\left(\frac{d}{d\rho}A(\rho)\right)\int_{\infty}^{\rho}d\xi\frac{\Theta_{>}(\xi)\Phi(\xi)A(\xi)}{F(\xi)}\nonumber \\
 &  & \qquad\qquad\quad-\int_{\infty}^{z}d\rho\frac{\Theta_{>}(\rho)}{r(\rho)^{2}}\left(\frac{d}{d\rho}A(\rho)\right)\int_{z_{H}}^{\rho}d\xi\frac{\Theta_{<}(\xi)\Phi(\xi)A(\xi)}{F(\xi)}\Bigg\}\:,\nonumber \\
a_{i}^{(13)}(z) & = & \bar{h}_{i}a^{(13)}(z)\\
 &  & +\epsilon_{ij}\bar{h}_{j}\frac{\lambda}{N_{r}}\left\{ \Theta_{>}(z)\int_{z_{H}}^{z}d\xi\Theta_{<}(\xi)S^{(13)}(\xi)-\Theta_{<}(z)\int_{\infty}^{z}d\xi\Theta_{>}(\xi)S^{(13)}(\xi)\right\} \:,\nonumber \\
h_{i}^{(14)}(z) & = & \bar{h}_{i}h^{(14)}(z)+\epsilon_{ij}\bar{h}_{j}\lambda\left\{ \int_{\infty}^{z}d\rho\frac{1}{r(\rho)^{2}F(\rho)}\left[\int_{z_{H}}^{\rho}d\xi S_{\textrm{Hall}}(\xi)+S_{\textrm{null}}(\rho)\right]\right\} \:,
\end{eqnarray}
with 
\begin{eqnarray}
S^{(13)}(\xi) & = & \frac{\Phi(\xi)A(\xi)}{F(\xi)}\int_{\infty}^{\xi}d\rho\frac{A(\rho)}{r(\rho)^{2}}\left(\frac{d}{d\rho}A(\rho)\right)+\frac{1}{N_{r}r(\xi)^{2}}\left[2F(\xi)\left(\frac{d}{d\xi}A(\xi)\right)^{2}+\lambda^{2}A(\xi)^{2}\right]\nonumber \\
 &  & \qquad\quad\times\left(\Theta_{<}(\xi)\int_{\infty}^{\xi}d\rho\Theta_{>}(\rho)\frac{\Phi(\rho)A(\rho)}{F(\rho)}-\Theta_{>}(\xi)\int_{z_{H}}^{\xi}d\rho\Theta_{<}(\rho)\frac{\Phi(\rho)A(\rho)}{F(\rho)}\right)\nonumber \\
 &  & +\frac{2\Phi(\xi)}{N_{r}F(\xi)}\Bigg\{\Theta_{<}(\xi)\int_{\infty}^{\xi}d\rho\Theta_{>}(\rho)\frac{A(\rho)}{r(\rho)^{2}}\left[F(\rho)\left(\frac{d}{d\rho}A(\rho)\right)^{2}+\lambda^{2}A(\rho)^{2}\right]\\
 &  & \qquad\quad-\Theta_{>}(\xi)\int_{z_{H}}^{\xi}d\rho\Theta_{<}(\rho)\frac{A(\rho)}{r(\rho)^{2}}\left[F(\rho)\left(\frac{d}{d\rho}A(\rho)\right)^{2}+\lambda^{2}A(\rho)^{2}\right]\Bigg\}\:,\nonumber \\
S_{\textrm{null}}(\rho) & = & F(\rho)\left(\frac{d}{d\rho}A(\rho)\right)\Bigg\{\frac{2}{N_{r}}\Bigg(\Theta_{<}(\rho)\int_{\infty}^{\rho}d\xi\Theta_{>}(\xi)S^{(13)}(\xi)\nonumber \\
 &  & \qquad-\Theta_{>}(\rho)\int_{z_{H}}^{\rho}d\xi\Theta_{<}(\xi)S^{(13)}(\xi)\Bigg)+A(\rho)S_{1}^{(14)}(\rho)\Bigg\}+S_{2}^{(14)}(\rho)\:,\\
S_{1}^{(14)}(\rho) & = & \frac{2}{N_{r}}\Bigg\{\int_{\infty}^{\rho}d\xi\frac{\Theta_{<}(\xi)}{r(\xi)^{2}}\left(\frac{d}{d\xi}A(\xi)\right)\int_{\infty}^{\xi}d\sigma\frac{\Theta_{>}(\sigma)\Phi(\sigma)A(\sigma)}{F(\sigma)}\nonumber \\
 &  & \qquad-\int_{\infty}^{\rho}d\xi\frac{\Theta_{>}(\xi)}{r(\xi)^{2}}\left(\frac{d}{d\xi}A(\xi)\right)\int_{z_{H}}^{\xi}d\sigma\frac{\Theta_{<}(\sigma)\Phi(\sigma)A(\sigma)}{F(\sigma)}\Bigg\}\:,\\
S_{2}^{(14)}(\rho) & = & \frac{2}{N_{r}}\Bigg\{\int_{z_{H}}^{\rho}d\xi\frac{\Theta_{<}(\xi)\Phi(\xi)A(\xi)}{F(\xi)}\int_{\infty}^{\rho}d\sigma\frac{\Theta_{>}(\sigma)A(\sigma)F(\sigma)}{r(\sigma)^{2}}\left(\frac{d}{d\sigma}A(\sigma)\right)^{2}\nonumber \\
 &  & \qquad-\int_{\infty}^{\rho}d\xi\frac{\Theta_{>}(\xi)\Phi(\xi)A(\xi)}{F(\xi)}\int_{z_{H}}^{\rho}d\sigma\frac{\Theta_{<}(\sigma)A(\sigma)F(\sigma)}{r(\sigma)^{2}}\left(\frac{d}{d\sigma}A(\sigma)\right)^{2}\Bigg\}\:,
\end{eqnarray}
and most importantly
\begin{eqnarray}
S_{\textrm{Hall}}(\xi) & = & \frac{2}{N_{r}}\frac{\Phi(\xi)A(\xi)}{F(\xi)}\Bigg\{\Theta_{>}(\xi)\int_{z_{H}}^{\xi}d\sigma\Theta_{<}(\sigma)\frac{A(\sigma)}{r(\sigma)^{2}}\left[2F(\sigma)\left(\frac{d}{d\sigma}A(\sigma)\right)^{2}+\lambda^{2}A(\sigma)^{2}\right]\nonumber \\
 &  & \qquad-\Theta_{<}(\xi)\int_{\infty}^{\xi}d\sigma\Theta_{>}(\sigma)\frac{A(\sigma)}{r(\sigma)^{2}}\left[2F(\sigma)\left(\frac{d}{d\sigma}A(\sigma)\right)^{2}+\lambda^{2}A(\sigma)^{2}\right]\Bigg\}\:.\label{S_Hall}
\end{eqnarray}
Our final results of correlators up to the desired order will not
contain $h^{(04)}(z)$, $h^{(14)}(z)$ and $a^{(13)}(z)$, so we will
not give their explicit expressions here. Notice that $S_{\textrm{null}}(z)\rightarrow0$
near the boundary $z\rightarrow\infty$. Because of this, we will
see later that $S_{\textrm{null}}(z)$ will drop off in the final
expression for Hall viscosity.

\subsection{2-Point Functions and Viscosities}

We follow the prescriptions in \cite{Herzog:2002pc,Barnes:2010jp,Arnold:2011ja}
to obtain causal 2-point functions from the above bulk-to-boundary
propagators and the second order on-shell action (\ref{S2_onshell}).
The total on-shell boundary action for the tensor mode is 
\begin{eqnarray}
S^{(2)} & = & \frac{1}{2\kappa^{2}}\int_{z=\infty}d^{3}x\sum_{i=e,o}\Bigg\{-\frac{1}{2}r(z)^{2}F(z)\left[h_{i}\left(\frac{d}{dz}h_{i}\right)\right]-F(z)\left[a_{i}\left(\frac{d}{dz}a_{i}\right)\right]\nonumber \\
 &  & \quad-\frac{1}{2}\left[\frac{d}{dz}\left(r(z)^{2}F(z)\right)-\frac{4}{R}r(z)^{2}\sqrt{F(z)}\right]h_{i}^{2}+F(z)\left(\frac{d}{dz}A(z)\right)h_{i}a_{i}\Bigg\}\:.\label{S2_tensor}
\end{eqnarray}
For a causal 2-point function, one of the two operators has earlier
time than the other one. The causal prescription is that in every
term in the above action, substitute one of the two fluctuation fields
with the advanced boundary-to-bulk propagator, whose boundary fields
are identified as the sources to the earlier-time operators, and substitute
the other fluctuation field with the retarded boundary-to-bulk propagator,
whose boundary fields are identified as sources to the later-time
operators.%
\footnote{Since every term is quadratic in fluctuations, there are two permutations
of substitution for each term.%
} The solution we obtain in the previous section using the incoming-wave
condition near the horizon, (\ref{h_propagator}) and (\ref{a_propagator}),
are the advanced boundary-to-bulk propagators. Notice that the equations
(\ref{EOM_h_tensor}) and (\ref{EOM_a_tensor}) are invariant under
the simultaneous reflections of $\omega\rightarrow-\omega$ and $\lambda\rightarrow-\lambda$,
the retarded boundary-to-bulk propagators can be obtained from the
advanced ones also by these operations. The readers should pay attention
to the operation $\lambda\rightarrow-\lambda$, which is very crucial
for getting the correct results. By taking functional derivatives
of the substituted boundary action (\ref{S2_tensor}) with respect
to the boundary fields $\bar{h}_{i}$, we get the causal 2-point functions:%
\footnote{The other two 2-point functions one can compute are not independent
from the above ones:
\[
G_{\textrm{ra}}^{xy,xx-yy}(\omega)=G_{\textrm{ra}}^{xx-yy,xy}(-\omega)\:,\qquad G_{\textrm{ra}}^{xx-yy,xx-yy}(\omega)=4G_{\textrm{ra}}^{xy,xy}(\omega)\:.
\]
} 
\begin{eqnarray}
G_{\textrm{ra}}^{xx-yy,xy}(\omega) & = & -i\omega\left(\frac{\lambda}{\kappa^{2}}\int_{z_{H}}^{\infty}dzS_{\textrm{Hall}}(z)\right)+O\left(\omega^{2},A(z)^{6}\right)\:,\\
G_{\textrm{ra}}^{xy,xy}(\omega) & = & -\frac{\Gamma}{2\kappa^{2}R^{2}}-i\omega\frac{r(z_{H})^{2}}{2\kappa^{2}}\left[1-2\int_{z_{H}}^{\infty}dz\frac{A(z)}{r(z)^{2}}\left(\frac{d}{dz}A(z)\right)\right]+O\left(\omega^{2},A(z)^{4}\right)\:,\nonumber \\
\end{eqnarray}
where $\Gamma$ is defined in (\ref{NB Conditions}). Compare with
Kubo formulae 
\begin{eqnarray}
G_{\textrm{ra}}^{xx-yy,xy}(\omega) & = & -2i\eta_{H}\omega+O\left(\omega^{2}\right)\:,\\
G_{\textrm{ra}}^{xy,xy}(\omega) & = & p-i\eta\omega+O\left(\omega^{2}\right)\:,
\end{eqnarray}
where $p\,(=\frac{1}{2}\varepsilon)$, $\eta$ and $\eta_{H}$ are
the pressure, shear viscosity and Hall viscosity of the system, and
also notice the entropy density $s$ given by (\ref{EntropyDensity}),
we have 
\begin{equation}
\eta_{H}=\frac{\lambda}{2\kappa^{2}}\int_{z_{H}}^{\infty}dzS_{\textrm{Hall}}(z)+O\left(A(z)^{6}\right)
\end{equation}
and 
\begin{equation}
\frac{\eta}{s}=\frac{1}{4\pi}\left\{ 1-2\int_{z_{H}}^{\infty}dz\frac{A(z)}{r(z)^{2}}\left(\frac{d}{dz}A(z)\right)+O\left(A(z)^{4}\right)\right\} 
\end{equation}
where $S_{\textrm{Hall}}(z)$ is given by (\ref{S_Hall}). Notice
that close to $T_{c}$, $A(z)$ is monotonic and vanishes at $z=\infty$,
which means the second term in the above ratio of $\eta/s$ is always
positive, thus the universal lower bound of $1/4\pi$ for $\eta/s$
is not violated up to this order. This is the question \cite{Natsuume:2010ky}
tried to address but failed, and we provide the answer now. For Canonical
ensemble, we get 
\begin{equation}
\frac{\eta}{s}=\frac{1}{4\pi}\left\{ 1+\frac{0.821}{(\lambda R)^{2}}\left[1-\frac{0.672}{(\lambda R)^{2}}+O\left(\frac{1}{(\lambda R)^{4}}\right)\right]\left(\frac{T_{c}-T}{T_{c}}\right)\theta\left(T_{c}-T\right)\right\} \:.
\end{equation}
For Grand Canonical ensemble 
\begin{equation}
\frac{\eta}{s}=\frac{1}{4\pi}\left\{ 1+\frac{0.701}{(\lambda R)^{2}}\left[1-\frac{0.486}{(\lambda R)^{2}}+O\left(\frac{1}{(\lambda R)^{4}}\right)\right]\left(\frac{T_{c}-T}{T_{c}}\right)\theta\left(T_{c}-T\right)\right\} \:.
\end{equation}
This ratio for anisotropic holographic $p$-wave superfluid models
has been computed in \cite{Erdmenger:2010xm,Erdmenger:2011tj,Erdmenger:2012zu}
and the behavior they found is similar to our result.

\subsection{Bulk-to-Bulk Propagators}

Now we will solve the two unknown functions $\Theta_{<}(z)$ and $\Theta_{>}(z)$
in the bulk-to-bulk propagators near the critical temperature and
obtain a more compact formula for Hall viscosity. Notice that (\ref{Theta_EOM})
is the same as (\ref{EQ3}) except for the non-linear $A(z)^{3}$
term. So we will use the same variational method to solve $\Theta(z)$
here, as we did in Section 4 for $A(z)$. 

Let us start with $\Theta_{<}(z)$. First, we make a perturbative
expansion: 
\begin{equation}
\Theta_{<}(z)=\Theta_{<}^{(1)}(z)\epsilon+\Theta_{<}^{(2)}(z)\epsilon^{3}+O\left(\epsilon^{5}\right)\:,
\end{equation}
and its near boundary value in (\ref{Theta_NB}) will get a similar
expansion: 
\[
B_{<}=B_{<}^{(1)}\epsilon+B_{<}^{(2)}\epsilon^{3}+O\left(\epsilon^{5}\right)\:.
\]
The function $F(z)$ and $\Phi(z)$ in (\ref{Theta_EOM}) have already
been expanded in (\ref{A_Expansion}) and computed in Section 4, so
we can just cite the results there. For the first order, the EOM that
$\Theta_{<}^{(1)}(z)$ satisfies is exactly the same as (\ref{EQ3_1}),
and we also impose the same near-horizon regularity condition for
them, so we can choose 
\begin{equation}
\Theta_{<}^{(1)}(z)=A^{(1)}(z)\:,
\end{equation}
then 
\begin{equation}
B_{<}^{(1)}=\alpha_{0}^{(1)}=c_{0}^{(1)}\alpha_{1}^{(1)}\:,
\end{equation}
where $c_{0}^{(1)}$ is given in (\ref{c0_1}) and $\alpha_{1}^{(1)}\propto\langle\mathcal{O}\rangle$.
Next we will solve for $\Theta_{<}^{(2)}$, similarly as we did for
$A^{(2)}(z)$, but not exactly the same, since their equations differ
by the non-linear $A^{(1)}(z)^{3}$ term. From now on we will define
\[
\Theta_{<}^{(2)}\equiv\tilde{A}^{(2)}(z)\:.
\]
The EOM for $\tilde{A}^{(2)}(z)$ is 
\begin{equation}
\frac{d}{dz}\left[F^{(0)}(z)\left(\frac{d}{dz}\tilde{A}^{(2)}(z)\right)\right]+\frac{\lambda^{2}\Phi^{(0)}(z)^{2}}{F^{(0)}(z)}\tilde{A}^{(2)}(z)+\tilde{\Upsilon}^{(2)}(z)=0\:,
\end{equation}
 where 
\begin{eqnarray}
\tilde{\Upsilon}^{(2)}(z) & = & \left(\frac{d}{dz}A^{(1)}(z)\right)\left[\left(\frac{d}{dz}F^{(1)}(z)\right)-\frac{F^{(1)}(z)}{F^{(0)}(z)}\left(\frac{d}{dz}F^{(0)}(z)\right)\right]\nonumber \\
 &  & -2\lambda^{2}A^{(1)}(z)\frac{\Phi^{(0)}(z)}{F^{(0)}(z)^{2}}\left(\Phi^{(0)}(z)F^{(1)}(z)-F^{(0)}(z)\Phi^{(1)}(z)\right)\:.
\end{eqnarray}
Comparing the above expression with $\Upsilon^{(2)}(z)$ in Section
4, the only difference is the $A^{(1)}(z)^{3}$ term. The action associated
with the variational problem is 
\begin{equation}
\tilde{I}^{(2)}=\int_{z_{H}}^{\infty}dz\left[-F^{(0)}(z)\left(\frac{d}{dz}\tilde{A}^{(2)}(z)\right)^{2}+\frac{\lambda^{2}\Phi^{(0)}(z)^{2}}{F^{(0)}(z)}\tilde{A}^{(2)}(z)^{2}+2\tilde{\Upsilon}^{(2)}(z)\tilde{A}^{(2)}(z)\right]+\tilde{I}_{\textrm{boundary}}^{(2)}\:.
\end{equation}
Following Section 4, we impose the same boundary conditions for $\tilde{A}^{(2)}(z)$:
one is the regularity condition near the horizon, and the other is
\begin{equation}
\tilde{\alpha}_{1}^{(2)}=0\:,
\end{equation}
where $\tilde{\alpha}_{1}^{(2)}$ is defined in the $\tilde{\;}$
version of (\ref{A2_NB}). We also have 
\begin{equation}
\tilde{I}_{\textrm{boundary}}^{(2)}=0\:.
\end{equation}
 The trial ansatz is 
\begin{equation}
\tilde{A}^{(2)}(z)=\left(1+\frac{z_{H}}{z}+\frac{z_{H}^{2}}{z^{2}}-\frac{q^{2}z_{H}^{3}}{4\lambda^{2}R^{2}z^{3}}\right)\left(\tilde{\alpha}_{0}^{(2)}-\frac{z_{H}\tilde{\alpha}_{0}^{(2)}}{z}+\sum_{i=2}^{n}\frac{\tilde{c}_{i}^{(2)}}{z^{2}}\right)
\end{equation}
 and we can solve $\tilde{\alpha}_{0}^{(2)}$ and $\tilde{c}_{i}^{(2)}$
in a similar fashion as before. We get $\tilde{\alpha}_{0}^{(2)}\propto\left(\alpha_{1}^{(1)}\right)^{3}$.
Define $\tilde{\alpha}_{0}^{(2)}=\tilde{c}_{0}^{(2)}\alpha_{1}^{(1)}$,
then $\tilde{c}_{0}^{(2)}\propto\left(\alpha_{1}^{(1)}\right)^{2}\propto\langle\mathcal{O}\rangle^{2}$.
We then have 
\begin{equation}
B_{<}^{(2)}=\tilde{\alpha}_{0}^{(2)}=\tilde{c}_{0}^{(2)}\alpha_{1}^{(1)}
\end{equation}
and 
\begin{equation}
B_{<}=B_{<}^{(1)}+B_{<}^{(2)}=\alpha_{0}^{(1)}+\tilde{\alpha}_{0}^{(2)}=\left(c_{0}^{(1)}+\tilde{c}_{0}^{(2)}\right)\alpha_{1}^{(1)}\:.
\end{equation}
Notice that near the critical line, both $c_{0}^{(1)}\propto\tilde{c}_{0}^{(2)}\propto\left(T_{c}-T\right)$,
so $B_{<}^{(1)}$ are in fact of the same order as $B_{<}^{(2)}$!
The calculation gives 
\begin{eqnarray}
c_{0}^{(1)}+\tilde{c}_{0}^{(2)} & = & \frac{1}{\kappa R^{\frac{3}{2}}\sqrt{\rho\lambda}}\left[-0.603+\frac{0.366}{(\lambda R)^{2}}+O\left(\frac{1}{(\lambda R)^{4}}\right)\right]\left(1-\frac{T}{T_{c}}\right)\theta\left(T_{c}-T\right)\nonumber \\
 & = & \frac{1}{\mu\lambda R^{3}}\left[-0.495+\frac{0.253}{(\lambda R)^{2}}+O\left(\frac{1}{(\lambda R)^{4}}\right)\right]\left(1-\frac{T}{T_{c}}\right)\theta\left(T_{c}-T\right)\:,\label{c0_12}
\end{eqnarray}
where the first line is for Canonical ensemble and second line Grand
Canonical ensemble. Notice that with the cubic term $A(z)^{3}$ in
place, we will get $\alpha_{0}^{(1)}+\alpha_{0}^{(2)}=c_{0}^{(1)}+c_{0}^{(2)}=0$,
which is exactly how the sourceless condition (\ref{NoSource_2})
is realized. Another comment we would like to make is regarding the
gauge freedom associated with $\Phi^{(1)}(z)$ which is discussed
in Section 4: only the combination $c_{0}^{(1)}+\tilde{c}_{0}^{(2)}$
is gauge-invariant (i.e. independent of the coefficient of that pure
gauge solution); both $c_{0}^{(1)}$ and $\tilde{c}_{0}^{(2)}$ alone
are not gauge-invariant when written in terms of \emph{physical} variables
like $T$ near the critical line.

Next we will solve $\Theta_{>}(z)$, but this is much easier. Using
(\ref{Nr_const}), we have the Wronskian $W_{r}\left[\Theta_{<}(z),\Theta_{>}(z)\right]\propto B_{<}\propto c_{0}^{(1)}+\tilde{c}_{0}^{(2)}\propto T_{c}-T$,
which vanishes at $T=T_{c}$. Thus at the critical line, $\Theta_{<}(z)$
and $\Theta_{>}(z)$ are not linearly independent: they are just proportional
to each other. Thus at the leading order, we can choose $\Theta_{>}^{(1)}(z)=\Theta_{<}^{(1)}(z)=A^{(1)}(z)$.
From now on for simplicity we will omit the superscript ``$^{(1)}$''
in $A^{(1)}(z)$ when there is no confusion, since throughout this
paper we always talk about near critical regime. Thus we have 
\begin{equation}
\Theta_{>}(z)=\Theta_{<}(z)=A(z)\:.
\end{equation}
From (\ref{Theta_NB}) we have 
\[
B_{>}=\alpha_{1}^{(1)}\:,
\]
then from (\ref{Nr_const}) the normalization constant of the bulk
Green's function is 
\begin{equation}
N_{r}=-\frac{1}{R^{2}}\left(c_{0}^{(1)}+\tilde{c}_{0}^{(2)}\right)\left(\alpha_{1}^{(1)}\right)^{2}\:,\label{Nr_nc}
\end{equation}
where $c_{0}^{(1)}+\tilde{c}_{0}^{(2)}$ is given in (\ref{c0_12}).

\subsection{Hall Viscosity}

Now the two terms in (\ref{S_Hall}) can be combined and then factored
out, which gives a simpler expression for Hall viscosity near the
critical temperature: 
\begin{equation}
\eta_{H}=\left(\frac{\lambda}{\kappa^{2}}\int_{z_{H}}^{\infty}d\xi\frac{\Phi(\xi)A(\xi)^{2}}{F(\xi)}\right)\left\{ \frac{1}{N_{r}}\int_{z_{H}}^{\infty}d\sigma\frac{A(\sigma)^{2}}{r(\sigma)^{2}}\left[2F(\sigma)\left(\frac{d}{d\sigma}A(\sigma)\right)^{2}+\lambda^{2}A(\sigma)^{2}\right]\right\} \:.\label{HallViscosity}
\end{equation}
We can see the above expression is factorized into two parts. To compute
the above expression, we set $A(z)=A^{(1)}(z)$, $r(z)=r^{(0)}(z)$,
$F(z)=F^{(0)}(z)$, $\Phi(z)=\Phi^{(0)}(z)$ and use the analytic
solutions obtained via variational method and $N_{r}$ given by (\ref{Nr_nc}).
For both ensembles, the second complicated factor inside ``$\left\{ \:\right\} $''
turns out to be 
\[
1.02-\frac{0.025}{(\lambda R)^{2}}+O\left(\frac{1}{(\lambda R)^{4}}\right)
\]
along the critical line, which is always close to $1$, since the
sub-leading orders give negligible corrections, even for small $\lambda R$.
Finally, for Canonical ensemble, we have 
\begin{eqnarray}
\frac{\eta_{H}}{\rho} & = & \frac{1.09}{\lambda R}\left[1-\frac{0.68}{(\lambda R)^{2}}+O\left(\frac{1}{(\lambda R)^{4}}\right)\right]\left(1-\frac{T}{T_{c}}\right)\theta\left(T_{c}-T\right)\:,\\
\frac{\eta_{H}}{s} & = & \frac{0.639}{(\lambda R)^{2}}\left[1-\frac{0.86}{(\lambda R)^{2}}+O\left(\frac{1}{(\lambda R)^{4}}\right)\right]\left(1-\frac{T}{T_{c}}\right)\theta\left(T_{c}-T\right)\:.
\end{eqnarray}
For Grand Canonical ensemble, we have 
\begin{eqnarray}
\frac{\eta_{H}}{\mu^{2}} & = & 1.01\left[1-\frac{0.32}{(\lambda R)^{2}}+O\left(\frac{1}{(\lambda R)^{4}}\right)\right]\frac{R^{2}}{\kappa^{2}}\left(1-\frac{T}{T_{c}}\right)\theta\left(T_{c}-T\right)\:,\\
\frac{\eta_{H}}{s} & = & \frac{0.545}{(\lambda R)^{2}}\left[1-\frac{0.67}{(\lambda R)^{2}}+O\left(\frac{1}{(\lambda R)^{4}}\right)\right]\left(1-\frac{T}{T_{c}}\right)\theta\left(T_{c}-T\right)\:.
\end{eqnarray}

\bigskip{}

\section{Vector Mode Fluctuations and Angular Momentum}

\subsection{Edge Current and Angular Momentum Density}

In this section we only study the static case, so there is no $t$-dependence
anywhere, and $\omega=0$. For fluids in $(2+1)$-dimensional flat
Minkowskian space, the $t$-component of the equation of energy-stress
tensor conservation reads 
\begin{equation}
\partial_{i}T^{ti}(\vec{x})=0
\end{equation}
and has the solution 
\begin{equation}
T^{ti}(\vec{x})=\epsilon^{ij}\partial_{j}\vartheta(\vec{x})\:,\label{T_ti_1}
\end{equation}
where $i,j,k=x,y$, $\vartheta(\vec{x})$ is an arbitrary function
and the totally anti-symmetric tensor $\epsilon^{ij}$ in flat $2$-dimensional
Euclidean space is normalized to $\epsilon^{xy}=1$. To proceed, we
put the fluid in a box of size $b$ (the shape does not really matter,
even though we assume it is a square) and at the end we can send $b\rightarrow\infty$.
If the fluid is almost homogeneous in the box, then to the leading
order in derivative expansion, we have 
\begin{equation}
\vartheta(\vec{x})=\begin{cases}
\frac{1}{2}\ell & \qquad(|x|\leqslant\frac{1}{2}b,\,|y|\leqslant\frac{1}{2}b)\\
0 & \qquad(\textrm{otherwise})
\end{cases}\:,\label{T_ti_2}
\end{equation}
where $\ell$ is a constant. So we can see $T^{ti}(\vec{x})$ is vanishing
both inside and outside the fluid (box), and is only non-vanishing
at the boundary of the fluid: 
\begin{equation}
T^{ti}(\vec{x})=\frac{1}{2}\ell\epsilon^{ij}\left[-\delta(x^{j}-\frac{1}{2}b)+\delta(x^{i}+\frac{1}{2}b)\right]\theta(\frac{1}{2}b-|x|)\theta(\frac{1}{2}b-|y|)\:.
\end{equation}
This is a momentum flow around the edge of the fluid -- the edge current,
and $\ell$ characterizes its strength. The direction of the edge
current is always along the edge, either clockwise or counter-clockwise,
depending on the sign of $\ell$. Notice that here for the uniform
hydrodynamic limit, $T^{ti}(\vec{x})$ at the leading order is vanishing
everywhere inside the bulk of the fluid, and this ``boundary'' nature
must be related to the topological nature of the underlying field
theory, as is commonly recognized in the study of Hall effect and
other parity-breaking effects. This ``boundary'' nature, also appearing
in its holographic dual theory, requires us to handle the boundary
terms with extreme care, as will be shown in the next subsection.
There were previous reports about failures of finding angular momenta
which were expected to be non-vanishing, and we suspect that the mishandling
of boundary terms could be a potential cause of the failures. In this
section we will show how to handle the boundary terms correctly to
get expected results.

Since $T^{ti}$ is the momentum density, the total angular momentum
$L$ of the fluid can be defined in the usual way: 
\begin{equation}
L=\int d^{2}\vec{x}\epsilon_{ij}x^{i}T^{tj}(\vec{x})\:.
\end{equation}
Use (\ref{T_ti_1}), (\ref{T_ti_2}) and $\epsilon_{ij}\epsilon^{jk}=-\delta_{i}^{k}$,
then integrate by parts, we have 
\begin{equation}
L=\int d^{2}\vec{x}\vartheta(\vec{x})\partial_{i}x^{i}=\ell\int_{|x|,|y|\leqslant\frac{1}{2}b}d^{2}\vec{x}=\ell V_{2}\:,
\end{equation}
where $V_{2}=b^{2}$ is the volume of the fluid (box). We can see
that $\ell$ is the angular momentum density of the fluid.

\subsection{Linear On-Shell Action}

On the other hand, the 1st order on-shell action linear to metric
fluctuations is the source term to the energy-stress tensor: 
\begin{equation}
S^{(1)}=\frac{1}{2}\int d^{3}x\bar{h}_{\mu\nu}(x)T^{\mu\nu}(x)\:.
\end{equation}
For the static case when we turn on only $\bar{h}_{ti}(\vec{x})$
sources and use the above results, we have 
\begin{equation}
S^{(1)}=\int d^{3}x\bar{h}_{ti}(\vec{x})T^{ti}(\vec{x})=\int d^{3}x\bar{h}_{ti}(\vec{x})\epsilon^{ij}\partial_{j}\vartheta(\vec{x})\:.
\end{equation}
Integrate by parts and let $b\rightarrow\infty$, we have: 
\begin{equation}
S^{(1)}=\frac{\ell}{2}\int d^{3}x\epsilon^{ij}\partial_{i}\bar{h}_{tj}(\vec{x})\:.\label{S1_fluid}
\end{equation}
It is well known that in holography $S^{(1)}$ is an integral of total
derivatives, as shown above, and thus a boundary term, because background
EOMs make the bulk part vanish. The usual treatment is to say that
the boundary terms are vanishing at ``boundaries'' like $x,y=\frac{1}{2}b$
and only non-vanishing at the AdS boundary $z=\infty$. However, here
due to the edge current effect, we will not assume the boundary terms
vanishing at $x,y=\frac{1}{2}b$. Actually once we have obtained $S^{(1)}$
in the bulk, we will \emph{not} integrate out the total derivatives;
instead we will just keep it as a bulk integral over total derivatives,
and by comparing it with (\ref{S1_fluid}) we can read off $\ell$
directly from its coefficient. Notice that in (\ref{S1_fluid}) we
have ignored higher order derivative terms since we assume $\vartheta(\vec{x})$
is almost homogeneous inside the fluid, so it is essentially a hydrodynamic
expansion up to leading order in derivatives.

Now we calculate $S^{(1)}=S_{\textrm{bulk}}^{(1)}+S_{\textrm{GH}}^{(1)}+S_{\textrm{ct}}^{(1)}$
from the bulk action (\ref{S1_onshell}) and its associated boundary
terms in the background (\ref{BackgroundAnsatz}). There are four
different parts in $S^{(1)}$. The first part is the $z$-derivative
term in (\ref{S1_onshell}), which is a boundary term at the AdS boundary
$z=\infty$, together with the two other boundary terms $S_{\textrm{GH}}^{(1)}$
and $S_{\textrm{ct}}^{(1)}$. Its contribution is 
\begin{align}
 & \frac{1}{2\kappa^{2}}\int_{z=\infty}d^{3}x\Bigg\{2r(z)\left[F(z)\left(\frac{d}{dz}r(z)\right)-\frac{r(z)}{R}\sqrt{F(z)}\right]h_{t}^{t}(\vec{x},z)\nonumber \\
 & \qquad\qquad+\frac{1}{2}\left[\frac{d}{dz}\left(r(z)^{2}F(z)\right)-\frac{4}{R}r(z)^{2}\sqrt{F(z)}\right]\left[h_{x}^{x}(\vec{x},z)+h_{y}^{y}(\vec{x},z)\right]\\
 & \qquad\qquad+r(z)^{2}\left(\frac{d}{dz}\Phi(z)\right)a_{t}^{\mathbf{3}}(\vec{x},z)-F(z)\left(\frac{d}{dz}A(z)\right)\left[a_{x}^{\mathbf{1}}(\vec{x},z)+a_{y}^{\mathbf{2}}(\vec{x},z)\right]\Bigg\}\:.\nonumber 
\end{align}
Using (\ref{NB Conditions}) this is
\begin{equation}
\frac{1}{2\kappa^{2}}\int_{z=\infty}d^{3}x\Bigg\{\frac{\Gamma}{2R^{2}}\left[2\bar{h}_{t}^{t}(\vec{x})-\bar{h}_{x}^{x}(\vec{x})-\bar{h}_{y}^{y}(\vec{x})\right]-\frac{\Phi_{1}}{R^{2}}\bar{a}_{t}^{\mathbf{3}}(\vec{x})+\frac{\alpha_{1}}{R^{2}}\left[\bar{a}_{x}^{\mathbf{1}}(\vec{x})+\bar{a}_{y}^{\mathbf{2}}(\vec{x})\right]\Bigg\}\:.
\end{equation}
Since we only turn on $\bar{h}_{tx}(\vec{x})$ and $\bar{h}_{ty}(\vec{x})$
boundary fields for angular momentum, the above term has no contribution
to $S^{(1)}$. The second part is the $t$-derivative term in (\ref{S1_onshell}).
Since we are considering the static case where all fluctuations are
independent of $t$, it is zero. The last two parts are from the $x$-
and $y$-derivative terms in (\ref{S1_onshell}): that involving the
sound mode and tensor mode fluctuations 
\begin{align}
 & \frac{1}{2\kappa^{2}}\int d^{4}x\Bigg\{-\frac{1}{2}\left(\frac{\partial^{2}}{\partial x^{2}}+\frac{\partial^{2}}{\partial y^{2}}\right)\left[2h_{t}^{t}(\vec{x},z)+h_{x}^{x}(\vec{x},z)+h_{y}^{y}(\vec{x},z)\right]\nonumber \\
 & \qquad\qquad+\frac{1}{2}\left(\frac{\partial^{2}}{\partial x^{2}}-\frac{\partial^{2}}{\partial y^{2}}\right)\left[h_{x}^{x}(\vec{x},z)-h_{y}^{y}(\vec{x},z)\right]+2\frac{\partial^{2}}{\partial x\partial y}h_{y}^{x}(\vec{x},z)\Bigg\}
\end{align}
is quadratic in derivatives, thus is of higher order. So the only
relevant part is that involving the vector mode fluctuations 
\begin{eqnarray}
S^{(1)} & = & \frac{1}{2\kappa^{2}}\int d^{4}x\Bigg\{\frac{\lambda\Phi(z)A(z)}{F(z)}\left[-\left(\frac{\partial}{\partial x}a_{t}^{\mathbf{2}}(\vec{x},z)\right)+\left(\frac{\partial}{\partial y}a_{t}^{\mathbf{1}}(\vec{x},z)\right)\right]\nonumber \\
 &  & \qquad\qquad+\frac{\lambda A(z)^{2}}{r(z)^{2}}\left[-\left(\frac{\partial}{\partial x}a_{y}^{\mathbf{3}}(\vec{x},z)\right)+\left(\frac{\partial}{\partial y}a_{x}^{\mathbf{3}}(\vec{x},z)\right)\right]\Bigg\}\:.\label{S1_bulk}
\end{eqnarray}
Since this action is already linear in spatial derivatives (momentum),
to solve the bulk fields $a_{t}^{\mathbf{1}}$, $a_{t}^{\mathbf{2}}$,
$a_{x}^{\mathbf{3}}$ and $a_{y}^{\mathbf{3}}$ as a response to boundary
source $\bar{h}_{tx}$ and $\bar{h}_{ty}$ up to the leading order,
we only need to work in the zero momentum limit $\vec{k}=0$ and $\omega=0$,
which significantly simplifies the EOMs.

\subsection{Vector Mode EOMs}

We will work under the bulk gauge condition 
\begin{equation}
\begin{cases}
h_{\mu z}=0 & \qquad(\mu=t,x,y,z)\\
\: a_{z}^{\mathbf{I}}\:=0 & \qquad(\mathbf{I}=\mathbf{1},\mathbf{2},\mathbf{3})
\end{cases}\:.\label{GaugeCondition}
\end{equation}
We first make some redefinitions of the coordinates and fields. In
the rest of this section, unless otherwise stated, we will let $i,j=e,o$
exclusively.%
\footnote{The indices $i,j=e,o$ are just short-hand notations introduced to
make the equations look more compact. It is not necessary to think
them as some co-variant indices that are raised and lowered by some
metric. In other words, equations containing $i,j=e,o$ are just components
of some covariant equations and themselves not covariant in some $(e,o)$-space.
By definition, repeated indices of $i,j$ are summed over $e,o$.%
} Define 
\begin{align}
\frac{1}{2}\left(h_{tx}+h_{ty}\right)\equiv r(z)^{2}h_{t}^{e}\:,\quad\quad & \quad\quad\frac{1}{2}\left(h_{tx}-h_{ty}\right)\equiv r(z)^{2}h_{t}^{o}\:,\nonumber \\
\frac{1}{2}\left(a_{t}^{\mathbf{1}}+a_{t}^{\mathbf{2}}\right)\equiv a_{t}^{e}\:,\quad\quad & \quad\quad\frac{1}{2}\left(a_{t}^{\mathbf{1}}-a_{t}^{\mathbf{2}}\right)\equiv a_{t}^{o}\:,\\
\frac{1}{2}\left(a_{x}^{\mathbf{3}}+a_{y}^{\mathbf{3}}\right)\equiv a_{e}^{\mathbf{3}}\:,\quad\quad & \quad\quad\frac{1}{2}\left(a_{x}^{\mathbf{3}}-a_{y}^{\mathbf{3}}\right)\equiv a_{o}^{\mathbf{3}}\:.\nonumber 
\end{align}
Fields in the left (right) column in the above definitions are even
(odd) under the joint parity operations $x\leftrightarrow y$ and
$\mathbf{1}\leftrightarrow\mathbf{2}$. In the limit $\omega=0$ and
$\vec{k}=0$ these two groups decouple from each other. The independent
equations are 
\begin{eqnarray}
\frac{d}{dz}\left[r(z)^{4}\left(\frac{d}{dz}h_{t}^{i}(z)\right)\right]+r(z)^{2}\left(\frac{d}{dz}\Phi(z)\right)\left(\frac{d}{dz}a_{i}^{\mathbf{3}}(z)\right) & = & S_{t}^{(h)i}(z)\:,\\
r(z)^{2}\left(\frac{d}{dz}\Phi(z)\right)\left(\frac{d}{dz}h_{t}^{i}(z)\right)+\frac{d}{dz}\left[F(z)\left(\frac{d}{dz}a_{i}^{\mathbf{3}}(z)\right)\right] & = & S_{i}^{\mathbf{3}}(z)\:,\\
\frac{d}{dz}\left(\frac{a_{t}^{i}(z)}{\Phi(z)}\right) & = & S_{t}^{(a)i}(z)\:,
\end{eqnarray}
where 
\begin{eqnarray}
S_{t}^{(h)i}(z) & = & \left[r(z)^{2}\left(\frac{d}{dz}A(z)\right)^{2}+\frac{\lambda^{2}}{F(z)}A(z)^{4}\right]h_{t}^{i}(z)-\frac{\lambda^{2}\Phi(z)}{F(z)}A(z)^{2}a_{i}^{\mathbf{3}}(z)\nonumber \\
 &  & -\left[r(z)^{2}\left(\frac{d}{dz}A(z)\right)\left(\frac{d}{dz}a_{t}^{i}(z)\right)+\frac{\lambda^{2}}{F(z)}A(z)^{3}a_{t}^{i}(z)\right]\:,\\
S_{i}^{\mathbf{3}}(z) & = & -\frac{\lambda^{2}\Phi(z)}{F(z)}A(z)^{2}h_{t}^{i}(z)+\frac{\lambda^{2}}{r(z)^{2}}A(z)^{2}a_{i}^{\mathbf{3}}(z)+\frac{\lambda^{2}\Phi(z)}{F(z)}A(z)a_{t}^{i}(z)\:,\\
S_{t}^{(a)i}(z) & = & \left[\frac{d}{dz}\left(\frac{A(z)}{\Phi(z)}\right)\right]h_{t}^{i}(z)+\frac{F(z)}{r(z)^{2}\Phi(z)^{2}}\left[\left(\frac{d}{dz}A(z)\right)a_{i}^{\mathbf{3}}(z)-A(z)\left(\frac{d}{dz}a_{i}^{\mathbf{3}}(z)\right)\right]\:,\nonumber \\
\end{eqnarray}
and all other equations involving vector mode fields are linear combinations
of the above equations and their $z$-derivatives. Notice that all
sources $S_{\ldots}^{\ldots}(z)$ in the above equations contain $A(z)$.
So when we treat $A(z)$ perturbatively, all sources can also be treated
perturbatively, and at the leading order they all vanish.

\subsection{Boundary-to-Bulk Propagators}

From (\ref{S1_bulk}) we can see that the two terms in the first line
are proportional to $A(z)$ and those in the second line to $A(z)^{2}$,
thus to obtain the leading order result for $S^{(1)}$, which is of
order $A(z)^{2}$, we only need to solve $h_{t}^{i}$ and $a_{i}^{\mathbf{3}}$
to zeroth order in (independent of) $A(z)$ and $a_{t}^{i}$ to linear
order in $A(z)$. First we solve the zeroth order equations for $h_{t}^{i}$
and $a_{i}^{\mathbf{3}}$: 
\begin{eqnarray}
\frac{d}{dz}\left[r(z)^{4}\left(\frac{d}{dz}h_{t}^{i}(z)\right)\right]+r(z)^{2}\left(\frac{d}{dz}\Phi(z)\right)\left(\frac{d}{dz}a_{i}^{\mathbf{3}}(z)\right) & = & 0\:,\\
r(z)^{2}\left(\frac{d}{dz}\Phi(z)\right)\left(\frac{d}{dz}h_{t}^{i}(z)\right)+\frac{d}{dz}\left[F(z)\left(\frac{d}{dz}a_{i}^{\mathbf{3}}(z)\right)\right] & = & 0\:.
\end{eqnarray}
For each $i=e,o$ these are two coupled second order homogeneous ODEs,
so they have four independent solutions. Two solutions are trivial
to see: $h_{t}^{i}=\textrm{constant}$, $a_{i}^{\mathbf{3}}=0$ and
$h_{t}^{i}=0$, $a_{i}^{\mathbf{3}}=\textrm{constant}$. For the other
two solutions, using (\ref{NH Conditions}) to solve these equations
near the horizon, we find one of them contains $\ln(z-z_{H})$ in
$a_{i}^{\mathbf{3}}$ so it is dropped by the regularity requirement
near the horizon. The last independent solution is 
\begin{equation}
\left\{ \begin{aligned} & h_{t}^{i}(z)=\frac{F(z)}{r(z)^{2}}\times\textrm{constant}\\
 & a_{i}^{\mathbf{3}}(z)=-\Phi(z)\times\textrm{same constant}
\end{aligned}
\right.\:.\label{FourthSolution}
\end{equation}
This can be checked by using the background equations (\ref{EQ1})-(\ref{EQ4}).
In Appendix (B) we show that this solution is of a pure-gauge form
which can be obtained by a residual gauge transformation. We require
that $h_{t}^{i}(z)$ vanishes at the horizon. The same condition has
been used in \cite{Liu:2012zm,Liu:2013cha}. Bearing in mind that
we only turn on $\bar{h}_{ti}$ source%
\footnote{We view the boundary fields $\bar{h}_{tx}$ and $\bar{h}_{t}^{x}$
etc are living in the 3-dimensional flat Minkowskian space where the
fluid (field theory) system lives. That means the indices of these
boundary fields are raised and lowered by 3-dimensional flat Minkowskian
metric, not by the 4-dimensional bulk metric (\ref{BackgroundAnsatz}).%
}, we have at the leading order 
\begin{equation}
\left\{ \begin{aligned} & h_{t}^{i}(\vec{x},z)=\left(\frac{F(z)}{r(z)^{2}}+O\left(\vec{\partial},A(z)\right)\right)\bar{h}_{t}^{i}(\vec{x})\\
 & a_{i}^{\mathbf{3}}(\vec{x},z)=\left(\Phi_{0}-\Phi(z)+O\left(\vec{\partial},A(z)\right)\right)\bar{h}_{t}^{i}(\vec{x})
\end{aligned}
\right.\:.
\end{equation}

Next we solve for $a_{t}^{i}(z)$. At the leading order
\begin{equation}
\frac{d}{dz}\left(\frac{a_{t}^{i}(z)}{\Phi(z)}\right)=0\:,
\end{equation}
which has solution $a_{t}^{i}(z)\propto\Phi(z)$. This solution is
also dropped because $\bar{a}_{t}^{i}$ source is not turned on. At
the next order the source term becomes 
\begin{equation}
S_{t}^{(a)i}(z)=\frac{\Phi_{0}F(z)}{\Phi(z)^{2}r(z)^{2}}\left(\frac{d}{dz}A(z)\right)\bar{h}_{t}^{i}+O\left(A(z)^{2}\right)\:,
\end{equation}
Requiring that $a_{t}^{i}(z)\rightarrow0$ near the boundary, the
solution is 
\begin{equation}
a_{t}^{i}(\vec{x},z)=\left[\Phi_{0}\Phi(z)\int_{\infty}^{z}d\xi\frac{F(\xi)}{r(\xi)^{2}\Phi(\xi)^{2}}\left(\frac{d}{d\xi}A(\xi)\right)+O\left(\vec{\partial},A(z)^{2}\right)\right]\bar{h}_{t}^{i}(\vec{x})\:.
\end{equation}

\subsection{Angular Momentum Density and Ratio to Hall Viscosity}

From now on we restore the convention that $i,j=x,y$ used at the
beginning of this section. Plug in the above solutions into (\ref{S1_bulk}),
we have 
\begin{eqnarray*}
S^{(1)} & = & -\frac{\lambda}{2\kappa^{2}}\int d^{3}x\int_{z_{H}}^{\infty}dz\Bigg\{\Bigg[\Phi_{0}\frac{\Phi(z)^{2}A(z)}{F(z)}\int_{\infty}^{z}d\xi\frac{F(\xi)}{r(\xi)^{2}\Phi(\xi)^{2}}\left(\frac{d}{d\xi}A(\xi)\right)\\
 &  & \qquad+\left(\Phi_{0}-\Phi(z)\right)\frac{A(z)^{2}}{r(z)^{2}}\Bigg]\epsilon^{ij}\partial_{i}\bar{h}_{tj}(\vec{x})+O\left(\vec{\partial}^{2},A(z)^{3}\right)\Bigg\}\:.
\end{eqnarray*}
Use (\ref{EQ3}) to integrate by parts the first term, and then compare
with (\ref{S1_fluid}), we find the angular momentum density is 
\begin{equation}
\ell=-\frac{\lambda}{\kappa^{2}}\int_{z_{H}}^{\infty}dz\left[\frac{\Phi_{0}F(z)^{2}}{\lambda^{2}\Phi(z)^{2}r(z)^{2}}\left(\frac{d}{dz}A(z)\right)^{2}+\left(\Phi_{0}-\Phi(z)\right)\frac{A(z)^{2}}{r(z)^{2}}\right]+O\left(\vec{\partial},A(z)^{3}\right)\:,\label{AngMomDensity}
\end{equation}
where $\Phi_{0}=\Phi(z=\infty)$.

To compute the above expressions, again we set $A(z)=A^{(1)}(z)$,
$r(z)=r^{(0)}(z)$, $F(z)=F^{(0)}(z)$, $\Phi(z)=\Phi^{(0)}(z)$ and
use the analytic solutions obtained via variational method in the
previous two sections. For canonical ensemble, we have 
\begin{eqnarray}
\frac{\ell}{\rho} & = & -\frac{2.16}{\lambda R}\left[1-\frac{1.00}{(\lambda R)^{2}}+O\left(\frac{1}{(\lambda R)^{4}}\right)\right]\left(1-\frac{T}{T_{c}}\right)\theta\left(T_{c}-T\right)\:,\\
\frac{\ell}{s} & = & -\frac{1.27}{(\lambda R)^{2}}\left[1-\frac{1.18}{(\lambda R)^{2}}+O\left(\frac{1}{(\lambda R)^{4}}\right)\right]\left(1-\frac{T}{T_{c}}\right)\theta\left(T_{c}-T\right)\:.
\end{eqnarray}
For Grand Canonical ensemble, we have 
\begin{eqnarray}
\frac{\ell}{\mu^{2}} & = & -2.00\left[1-\frac{0.63}{(\lambda R)^{2}}+O\left(\frac{1}{(\lambda R)^{4}}\right)\right]\frac{R^{2}}{\kappa^{2}}\left(1-\frac{T}{T_{c}}\right)\theta\left(T_{c}-T\right)\:,\\
\frac{\ell}{s} & = & -\frac{1.08}{(\lambda R)^{2}}\left[1-\frac{0.99}{(\lambda R)^{2}}+O\left(\frac{1}{(\lambda R)^{4}}\right)\right]\left(1-\frac{T}{T_{c}}\right)\theta\left(T_{c}-T\right)\:.
\end{eqnarray}
In (\ref{eq:Condensate_Canon}) and (\ref{eq:Condensate_GrandCanon})
we have shown that the condensate $\langle\mathcal{O}\rangle\sim\sqrt{T_{c}-T}$,
which implies the superfluid density $n_{s}\sim\langle\mathcal{O}\rangle^{2}\sim T_{c}-T$.
This is the standard behavior one would expect from Ginzburg-Landau
theory. Here we also have $\ell\sim T_{c}-T$, which gives 
\[
\frac{\ell}{n_{s}}\sim\textrm{constant}\:.
\]
This can be understood as a statement that each Cooper pair possesses
a fixed amount of angular momentum. For a $p_{x}+ip_{y}$-wave Cooper
pair, this number shall be just $\hbar=1$. For us, the precise value
depends on the normalization constant in the relation $n_{s}\sim\langle\mathcal{O}\rangle^{2}$,
which we will not determine explicitly here. 

At the end, we find the ratio between Hall viscosity and angular momentum
density to be 
\begin{equation}
\frac{\eta_{H}}{\ell}=-0.504\left[1+\frac{0.32}{(\lambda R)^{2}}+O\left(\frac{1}{(\lambda R)^{4}}\right)\right]
\end{equation}
from both ensembles. The minus sign is also in agreement with \cite{Read:2008rn},
but it can differ if the angular momentum or Hall viscosity is defined
up to a sign. At large $\lambda R$, which corresponds to the probe
limit regime where back-reactions to the metric can be neglected,
the magnitude of this ratio is numerically $1/2$. As $\lambda R$
drops, the magnitude of the ratio increases. In the next section,
we will see that at low temperature near the critical $\lambda_{c}R\approx1$,
it diverges logarithmically as $\ln T$ following the same behavior
of Hall viscosity.

\bigskip{}

\section{Low Temperature Limit}

\subsection{About the Complete Phase Diagram}

In this section, we investigate the low temperature limit of the holographic
$p_{x}+ip_{y}$ model. This is also the small $\lambda$ limit along
the critical regime, since $T_{c}$ is a monotonic increasing function
of $\lambda$ as can be seen from the phase diagrams in Figure \ref{Fig:PhaseDiagrams}.
The corresponding gravity dual is the near-extremal limit of the AdS-RN
black hole (\ref{AdSRN Solution}) and its hairy brother, and at the
leading order, the extremal limit (\ref{Extremal Limit}) represents
$T\rightarrow0$ limit. This \emph{is} the low temperature limit of
the holographic $p_{x}+ip_{y}$ phase, but may not be the dual of
the low temperature limit of the actual field theory/condensed matter
systems, due to various instabilities. So before we start the computation
at the near-extremal limit, we would like to briefly comment on what
we do not consider here.

The phase diagrams of Figure \ref{Fig:PhaseDiagrams} may not be the
complete phase diagrams, because they are obtained by assuming that
the AdS-RN black hole and its hairy version discussed in Section 4
are the only two possible competing ground states of the system. At
low temperature this is usually not the case, but even so, there are
still complications. In Section 4, we have obtained that the characteristic
function differences between the two phases are quadratic in $T_{c}-T$,
and concluded that the phase transition along the critical line is
second order. But there is an assumption made implicitly to reach
such a conclusion, that is, the coefficient $a^{(2)}$ in (\ref{a_2})
is always positive. However, as can be seen from (\ref{a_2}), at
some small $\lambda R\sim2$, $a^{(2)}$ can become zero and then
negative as $\lambda$ decreases. The consequence is that $\langle\mathcal{O}\rangle$
as a function of $T$ becomes multivalued near $T_{c}$, thus the
phase transition becomes first order.%
\footnote{However, in the calculation for the near-extremal limit to be given
later in this section, we find $a^{(2)}$ is still positive, so the
phase transition is still second order there.%
} This phenomena has been observed in holographic $s$-wave models
\cite{Herzog:2008he,Basu:2008st} and $p$-wave models \cite{Ammon:2009xh,Gubser:2010dm}.
Here we see this can happen in $p_{x}+ip_{y}$ model as well. 

Another class of complications arises concerning the Hawking-Page
transition between pure AdS and AdS-black hole backgrounds and their
instabilities \cite{Witten:1998zw,Cvetic:1999ne,Chamblin:1999tk,Chamblin:1999hg,Mitra:1999ge,Gubser:2000mm}:
at low temperature, AdS-type backgrounds can be thermodynamically
favored over black hole backgrounds. This is also important for zero
temperature limit, because an extremal black hole has non-vanishing
horizon area thus non-zero entropy, which implies the ground state
is degenerate. But the real ground state shall be non-degenerate,
thus it can not be described by an extremal black hole, but an AdS-type
background with vanishing horizon area. In field theory language,
the phase transition between AdS and black hole is a confinement/deconfinement
transition, and in condensed matter language, insulator/conductor
transition. The low/zero temperature limit of holographic $s$-wave
models have been studied based on AdS domain wall geometries in \cite{Gubser:2008wz,Gubser:2008pf,Gubser:2009cg,Horowitz:2009ij,Konoplya:2009hv}
and on AdS solitons in \cite{Nishioka:2009zj,Horowitz:2010jq,Brihaye:2011vk},
and that of the anisotropic $p$-wave model based on AdS domain walls
in \cite{Basu:2009vv,Basu:2011np} and on AdS solitons in \cite{Akhavan:2010bf,Cai:2011ky}.
So far we have not seen any study on the isotropic $p_{x}+ip_{y}$
model, possibly because of the instability discussed in \cite{Gubser:2008wv}.

In the following, we will ignore all these complications. If the actual
low temperature state of the model can be a $p$-wave model which
breaks isotropy, then the hydrodynamic analysis presented at the beginning
of this paper, and the formula (\ref{eq:Kubo_HallVisc}), will be
invalid. The notion of Hall viscosity itself may not be even useful
or well-defined if isotropy is lost. The main purpose of studying
the near extremal limit of $p_{x}+ip_{y}$ model is to see how the
results we have presented in the previous sections, mostly as series
expansions for large $\lambda R$, can be extrapolated down to small
$\lambda R$ regime to give an overall qualitative picture for all
values of $\lambda R$. For this purpose, it is reasonable to ignore
all the complications mentioned above.

\subsection{Near-Extremal AdS-RN Black Hole with Condensate}

For the low temperature limit, if we directly work with the extremal
limit of AdS-RN black hole, we will encounter near-horizon divergence
for the angular momentum density. To understand its origin and how
to handle it properly, let us first have a look at the near-extremal
limit. The near-extremal limit of (\ref{AdSRN Solution}) corresponds
to take the following limit of the parameters: 
\begin{equation}
\left\{ \begin{aligned} & q=2\sqrt{3-6\tau}\lambda R\\
 & T=\frac{3z_{H}}{2\pi R^{2}}\tau
\end{aligned}
\right.\qquad\textrm{with}\qquad\tau\ll1\:.\label{Near Extremal Limit}
\end{equation}
Then the inner horizon is 
\begin{equation}
z_{h}=z_{H}\left(1-\tau\right)\:.
\end{equation}
The metric component $F^{(0)}(z)$ becomes 
\begin{equation}
F^{(0)}(z)=\frac{(z-z_{H})(z-z_{h})(z^{2}+2z_{H}z+3z_{H}^{2})}{z^{2}R^{2}}\:.
\end{equation}
Now if we look at (\ref{AngMomDensity}), near the horizon $z=z_{H}$
we encounter integrals like 
\[
\int\frac{dz}{z-z_{h}}=\ln(z-z_{h})\:.
\]
When we evaluate the integral at the lower bound $z=z_{H}$, we get
$\ln(z_{H}-z_{h})\sim\ln(\tau)$. So if we work directly with the
extremal black hole with $\tau=0$ from the very beginning, we will
encounter divergence.%
\footnote{Actually, $\ln(z-z_{h})$ has already appear earlier. If we go through
the same procedures as outlined in Section 4 to solve the model analytically,
we will get $\ln(z-z_{h})$ when we solve $r^{(1)}(z)$, $F^{(1)}(z)$
and $\Phi^{(1)}(z)$ using the integrals listed in Appendix (C), as
well as in $A^{(2)}(z)$. But except for $r^{(1)}(z)$, $\ln(z-z_{h})$
in the other functions is always multiplied by some factors of $z-z_{H}$
or $z-z_{h}$. So when evaluated at $z=z_{H}$, $\ln(\tau)$ will
always drop off or appear in higher order, thus if we start directly
with extremal black hole, we will not encounter divergence at these
intermediate steps, until we reach $\ell$ in (\ref{AngMomDensity}).%
} So we see $\tau$ serves as a near-horizon regulator. If we choose
to start with the extremal black hole, the solution will be that when
we encounter such a divergence near the horizon, instead of setting
the lower bound of the integral to be at $z=z_{H}$, we set it at
$z=z_{H}(1+\tau)$. At the leading order this strategy will give us
the same results as we work in near-extremal limit. This is what we
will do in the following.

\subsection{Extremal AdS-RN Black Hole with Condensate}

Now we can safely start with the extremal AdS-Reissner-Nordström Black
Hole 
\begin{equation}
\left\{ \begin{aligned} & r^{(0)}(z)=\frac{z}{R}\\
 & F^{(0)}(z)=\frac{(z-z_{H})^{2}(z^{2}+2z_{H}z+3z_{H}^{2})}{z^{2}R^{2}}\\
 & \Phi^{(0)}(z)=\frac{2\sqrt{3}z_{H}}{R}\left(1-\frac{z_{H}}{z}\right)
\end{aligned}
\right.\:,\label{AdSRN Extremal}
\end{equation}
then we will follow closely the same analytic procedure we have used
in the previous sections. To avoid redundancy and repetition, we will
only outline the differences here. The trial functions we use for
$A^{(i)}(z)$ are: 
\begin{eqnarray}
A^{(1)}(z) & = & \alpha_{1}^{(1)}\left(1+2\frac{z_{H}}{z}+3\frac{z_{H}^{2}}{z^{2}}\right)\left(c_{0}^{(1)}+\frac{1-2z_{H}c_{0}^{(1)}}{z}+\sum_{i=2}^{n}\frac{c_{i}^{(1)}}{z^{2}}\right)\:,\\
A^{(2)}(z) & = & \left(1+2\frac{z_{H}}{z}+3\frac{z_{H}^{2}}{z^{2}}\right)\left(\alpha_{0}^{(2)}-\frac{2z_{H}\alpha_{0}^{(2)}}{z}+\sum_{i=2}^{n}\frac{c_{i}^{(2)}}{z^{2}}\right)\:,
\end{eqnarray}
and $\tilde{A}^{(2)}(z)$ is just the $\tilde{\;}$ version of the
second line. From (\ref{Tc Condition}) we get the minimal coupling
that can trigger a phase transition, i.e. the critical coupling, is
\begin{equation}
\lambda_{c}R=0.745\:.
\end{equation}
Since the temperature is already very low, we will not parametrize
the deviation from the critical line as $T_{c}-T$, but instead as
$\lambda-\lambda_{c}$.

For Canonical ensemble, we obtain: 
\begin{eqnarray}
\frac{\langle\mathcal{O}\rangle}{\rho} & = & 0.939\left(\frac{\lambda}{\lambda_{c}}-1\right)^{\frac{1}{2}}\theta\left(\lambda-\lambda_{c}\right)\:,\\
c_{0}^{(1)}+\tilde{c}_{0}^{(2)} & = & -\frac{2.91}{\sqrt{\rho}R\kappa}\left(\frac{\lambda}{\lambda_{c}}-1\right)\theta\left(\lambda-\lambda_{c}\right)\:,
\end{eqnarray}
then
\begin{eqnarray}
\frac{\eta_{H}}{\rho} & = & -2.14\ln(\tau)\left(\frac{\lambda}{\lambda_{c}}-1\right)\theta\left(\lambda-\lambda_{c}\right)\:,\\
\frac{\eta_{H}}{s} & = & -1.18\ln(\tau)\left(\frac{\lambda}{\lambda_{c}}-1\right)\theta\left(\lambda-\lambda_{c}\right)\:,
\end{eqnarray}
and 
\begin{eqnarray}
\frac{\ell}{\rho} & = & -3.73\left(\frac{\lambda}{\lambda_{c}}-1\right)\theta\left(\lambda-\lambda_{c}\right)\:,\\
\frac{\ell}{s} & = & -2.06\left(\frac{\lambda}{\lambda_{c}}-1\right)\theta\left(\lambda-\lambda_{c}\right)\:,
\end{eqnarray}
with 
\[
\tau=\frac{T}{\sqrt{\hat{\rho}}}\:,
\]
and $\hat{\rho}=\frac{\kappa^{2}}{(2\pi)^{3}R^{2}}\rho$.

For Grand Canonical ensemble, we obtain: 
\begin{eqnarray}
\frac{\langle\mathcal{\hat{O}}\rangle}{\mu^{2}} & = & 4.37\times10^{-3}\left(\frac{\lambda}{\lambda_{c}}-1\right)^{\frac{1}{2}}\theta\left(\lambda-\lambda_{c}\right)\:,\\
c_{0}^{(1)}+\tilde{c}_{0}^{(2)} & = & -\frac{2.71}{\mu R^{2}}\left(\frac{\lambda}{\lambda_{c}}-1\right)\theta\left(\lambda-\lambda_{c}\right)\:,
\end{eqnarray}
then
\begin{eqnarray}
\frac{\eta_{H}}{\mu^{2}} & = & -2.47\ln(\tau)\left(\frac{\lambda}{\lambda_{c}}-1\right)\theta\left(\lambda-\lambda_{c}\right)\:,\\
\frac{\eta_{H}}{s} & = & -1.18\ln(\tau)\left(\frac{\lambda}{\lambda_{c}}-1\right)\theta\left(\lambda-\lambda_{c}\right)\:,
\end{eqnarray}
and 
\begin{eqnarray}
\frac{\ell}{\mu^{2}} & = & -4.30\frac{R^{2}}{\kappa^{2}}\left(\frac{\lambda}{\lambda_{c}}-1\right)\theta\left(\lambda-\lambda_{c}\right)\:,\\
\frac{\ell}{s} & = & -2.06\left(\frac{\lambda}{\lambda_{c}}-1\right)\theta\left(\lambda-\lambda_{c}\right)\:,
\end{eqnarray}
with 
\[
\tau=\frac{T}{\mu}\:,
\]
and $\hat{\mathcal{O}}=\frac{\kappa^{2}}{(2\pi)^{3}R^{2}}\mathcal{O}$.

For both ensembles, we have 
\begin{eqnarray}
\frac{\eta_{H}}{l} & = & 0.573\ln(\tau)\:,\\
\frac{\eta}{s} & = & \frac{1}{4\pi}\left\{ 1+11.7\left(\frac{\lambda}{\lambda_{c}}-1\right)\theta\left(\lambda-\lambda_{c}\right)\right\} \:.
\end{eqnarray}

\bigskip{}

\section{Conclusions and Comments}

In this paper we have studied the spontaneous parity breaking effect
of the holographic $p_{x}+ip_{y}$ model of \cite{Gubser:2008zu}.
We have proposed an analytic approach to solve such holographic models
by taking into full consideration of back-reactions. The method we
have shown here for computing the angular momentum density of the
model is general and can be used for other holographic models as well.
We obtain analytic expressions for Hall viscosity and angular momentum
density near the critical regime and find that the relation (\ref{HL_Relation})
between them holds in the probe limit regime where back-reations to
metric can be ignored. The effect of angular momentum density is to
accumulate momentum on the edge of the fluid system. Thus an edge
current of momentum is generated, and its intensity is proportional
to the angular momentum density. 

At the end we would like to make several comments on the results and
the holographic $p_{x}+ip_{y}$ model we use in this paper:
\begin{enumerate}
\item The near-critical behavior of Hall viscosity we have found in $p$-wave
model $\eta_{H}\propto\left(T_{c}-T\right)$, which is different from
that found in gravitational Chern-Simons model \cite{Saremi:2011ab,Chen:2011fs}
where $\eta_{H}\propto\sqrt{T_{c}-T}$. The reason is that in the
former case the condensate $A(z)$ enters the EOMs and thus the final
results quadratically, while in the latter the condensate -- the axion
scalar -- enters linearly. 
\item We have only studied the near-critical regime of the model, because
only this regime can be computed by analytic approaches. It is natural
to ask how Hall viscosity and angular momentum density behave deep
inside the superconducting phase. The complication is that a finite
non-vanishing $A(z)$ spoils the integrability of EOMs for the tensor
mode fluctuations (\ref{EOM_h_tensor}) and (\ref{EOM_a_tensor}),
thus it is hard to find a closed form for Hall viscosity written in
term of $A(z)$ and others. Going deep inside the symmetry-breaking
phase requires numeric techniques, so we will leave this to future
exploration. 
\item As can be seen from (\ref{HallViscosity}) and (\ref{AngMomDensity}),
results for Hall viscosity and angular momentum density do not have
the feature of membrane paradigm, with which such quantities can be
written solely in terms of near-horizon fields, as found in \cite{Saremi:2011ab}.
Our results are written as integrals of the condensate over the whole
region outside the black hole horizon, up to the boundary. This means
the IR degrees of freedom interact non-trivially with UV degrees of
freedom.
\item It is well known that the physics of Hall conductivity can be described
at the low energy effective theory level by a gauge Chern-Simons term
and is related to the topological nature of the states. This is also
the spirit of many holographic constructions. So is true for Hall
viscosity, which can be describe by a Wen-Zee term \cite{Hoyos:2011ez,Son:2013rqa}
in the presence of external magnetic field or a Gravitational Chern-Simons
term \cite{Saremi:2011ab} for pure thermal cases. However, in the
holographic $p_{x}+ip_{y}$ model, there is no Chern-Simons term in
the action and we do not see the topological structure at this level.
It will be interesting to see how the Chern-Simons term can be induced
at the parity-breaking ground state and what the topological structure
looks like (if there is any). This may possibly shed some light on
the second comments above, and on understanding the relation of holographic
$p_{x}+ip_{y}$ model to topological quantum states.
\end{enumerate}
It is also interesting to see whether other holographic models with
(either explicitly or spontaneously) broken parity respect (\ref{HL_Relation}),
once the angular momentum density is correctly computed. The gravitational
Chern-Simons model studied in \cite{Saremi:2011ab,Chen:2011fs} has
recently been shown in \cite{Wu:2013vya} to possess an angular momentum
density. There, Hall viscosity, angular momentum density and their
ratio all have complicated behaviors (numerically, even though the
analytic expressions look simple) below and near the critical temperature
and far away from the relation (\ref{HL_Relation}). A comprehensive
understanding of the generation of Hall viscosity and angular momentum
in generic holographic models, particularly for the gapped phases,
still requires future works.

\bigskip{}

\section*{Acknowledgments\textmd{\normalsize{ \addcontentsline{toc}{section}{Acknowledgments}}}}

We thank Wan-Zhe Feng, Yan He, Kathryn Levin, Hong Liu, Dung Xuan
Nguyen, Hirosi Ooguri, Matthew Roberts, Gordon Semenoff, Misha Stephanov,
Bogdan Stoica and Paul Wiegmann for useful discussions and comments.
This work is supported, in part, by DOE grant DE-FG02-90ER-40560,
NSF DMS-1206648 and a Simons Investigator grant from the Simons Foundation.

\bigskip{}

\bigskip{}

\begin{appendices}

\section{Green's Function }

We start with the general form of a linear second order inhomogeneous
ODE 
\begin{equation}
\frac{d}{dz}\left[P(z)\left(\frac{d}{dz}\phi(z)\right)\right]+Q(z)\phi(z)=S(z)\:,
\end{equation}
where $z\in[a,b]$ and $P(z)$, $Q(z)$ and $S(z)$ are known functions
of $z$. We assume $\Theta_{m}(z)$ ($m=<,>$) are two independent
solutions to the homogeneous equation 
\begin{equation}
\frac{d}{dz}\left[P(z)\left(\frac{d}{dz}\Theta_{m}(z)\right)\right]+Q(z)\Theta_{m}(z)=0\label{HomoEq}
\end{equation}
and satisfy appropriate boundary conditions at the two boundaries:
\begin{equation}
\Theta_{<}(z){\Big|_{z\rightarrow a}}=A\left(z-a\right)^{\alpha}\:,\qquad\Theta_{>}(z){\Big|_{z\rightarrow b}}=B\left(z-b\right)^{\beta}\:.
\end{equation}
The Green's function $G(z,z')$ satisfies similar boundary conditions
and 
\begin{equation}
\frac{d}{dz}\left[P(z)\left(\frac{d}{dz}G(z,z')\right)\right]+Q(z)G(z,z')=\delta(z-z')\:,\label{GreenEq}
\end{equation}
where $\delta(z-z')$ is the Dirac delta function. Then we have 
\begin{equation}
\phi(z)=\int_{a}^{b}dz'G(z,z')S(z')\:.
\end{equation}
The Green's function can be written in terms of the above two independent
solutions: 
\begin{equation}
G(z,z')=\frac{1}{N_{r}}\left\{ \Theta_{<}(z)\Theta_{>}(z')\theta(z'-z)+\Theta_{<}(z')\Theta_{>}(z)\theta(z-z')\right\} \:,
\end{equation}
and the normalization constant $N_{r}$ can be calculated by integrating
(\ref{GreenEq}) from $z=z'_{-}$ to $z=z'_{+}$, which gives 
\begin{equation}
N_{r}=P(z)\textrm{Wr}(z)\:,
\end{equation}
where the Wronskian is 
\begin{equation}
\textrm{Wr}(z)\equiv\Theta_{<}(z)\left(\frac{d}{dz}\Theta_{>}(z)\right)-\Theta_{>}(z)\left(\frac{d}{dz}\Theta_{<}(z)\right)\:.
\end{equation}
One can easily show from (\ref{HomoEq}) that $P(z)\textrm{Wr}(z)$,
even though it's a product of two functions of $z$, is indeed a constant
independent of $z$, thus $N_{r}=\textrm{constant}$. Assume near
the boundary $z=b$: 
\begin{equation}
\begin{cases}
P(z)\rightarrow P_{0}(z-b)^{\gamma}\\
\Theta_{<}(z)\rightarrow B_{<}(z-b)^{\beta_{<}}\\
\Theta_{>}(z)\rightarrow B_{>}(z-b)^{\beta_{>}}
\end{cases}\:,
\end{equation}
and $Q(z)$ are sub-leading to $P(z)$. The indicial equation of (\ref{HomoEq})
gives 
\begin{eqnarray}
\beta_{<},\;\beta_{>} & = & 0\quad\textrm{or}\quad1-\gamma\:,\\
\beta_{<}+\beta_{>} & = & 1-\gamma\:,
\end{eqnarray}
thus 
\begin{equation}
N_{r}=\left(\beta_{>}-\beta_{<}\right)P_{0}B_{<}B_{>}\:.
\end{equation}
The solution to the inhomogeneous equation is 
\begin{equation}
\phi(z)=\frac{1}{N_{r}}\left\{ \Theta_{<}(z)\int_{z}^{b}dz'\Theta_{>}(z')S(z')+\Theta_{>}(z)\int_{a}^{z}dz'\Theta_{<}(z')S(z')\right\} \:.\label{SolutionbyGreen}
\end{equation}
This is the formula we will use in the text to solve (\ref{EOM_a_tensor}).
Furthermore, one can integrate by parts the above expression to get
an alternative version 
\begin{eqnarray}
\phi(z) & = & \frac{1}{N_{r}}\Bigg\{\Theta_{<}(z)\int_{b}^{z}dz'\left(\frac{d}{dz'}\Theta_{>}(z')\right)\int_{a}^{z'}dz''S(z'')\\
 &  & -\Theta_{>}(z)\int_{a}^{z}dz'\left(\frac{d}{dz'}\Theta_{<}(z')\right)\int_{a}^{z'}dz''S(z'')+\Theta_{>}(b)\Theta_{<}(z)\int_{a}^{b}dz'S(z')\Bigg\}\:.\nonumber 
\end{eqnarray}
We can see that when one $\Theta_{m}(z)$ is constant (when $Q(z)=0$,
such as in (\ref{EOM_h_tensor})) this expression gives immediately
the result as one obtains by directly integrating the equation.

\bigskip{}

\section{Residual Gauge Transformations }

The gauge condition (\ref{GaugeCondition}) does not completely fix
the gauge. There are still residual gauge freedoms in the other non-vanishing
components of $h_{\mu\nu}$ and $a_{\mu}^{\mathbf{I}}$. These remaining
gauge freedoms are not strong enough to set any of these fields to
zero, but they can be used to gauge away certain parts of them --
the pure gauge solutions. The gauge transformations for Einstein-$SU(2)$
system are 
\begin{eqnarray}
\delta g_{\mu\nu} & = & -\nabla_{\mu}\xi_{\nu}-\nabla_{\nu}\xi_{\mu}\:,\\
\delta A_{\mu}^{\mathbf{I}} & = & -\xi^{\nu}\nabla_{\nu}A_{\mu}^{\mathbf{I}}+A^{\mathbf{I}\nu}\nabla_{\mu}\xi_{\nu}-\nabla_{\mu}\Lambda^{\mathbf{I}}+\epsilon^{\mathbf{IJK}}A_{\mu}^{\mathbf{J}}\Lambda^{\mathbf{K}}\:,
\end{eqnarray}
where $\xi^{\mu}$ and $\Lambda^{\mathbf{I}}$ are the gauge parameters.
Here we only consider the static case, where $\xi^{\mu}=\xi^{\mu}(x,y,z)$
and $\Lambda^{\mathbf{I}}=\Lambda^{\mathbf{I}}(x,y,z)$. The residual
gauge transformations need to preserve the gauge condition (\ref{GaugeCondition}),
which means 
\begin{eqnarray}
0 & = & \delta g_{tz}=F(z)\left(\frac{\partial}{\partial z}\xi^{t}\right)\:,\\
0 & = & \delta g_{xz}=-r(z)^{2}\left(\frac{\partial}{\partial z}\xi^{x}\right)-\frac{1}{F(z)}\left(\frac{\partial}{\partial x}\xi^{z}\right)\:,\\
0 & = & \delta g_{yz}=-r(z)^{2}\left(\frac{\partial}{\partial z}\xi^{y}\right)-\frac{1}{F(z)}\left(\frac{\partial}{\partial y}\xi^{z}\right)\:,\\
0 & = & \delta g_{zz}=\frac{1}{F(z)^{2}}\left[\left(\frac{d}{dz}F(z)\right)\xi^{z}-2F(z)\left(\frac{\partial}{\partial z}\xi^{z}\right)\right]\:,\\
0 & = & \delta A_{z}^{\mathbf{1}}=-A(z)\left(\frac{\partial}{\partial z}\xi^{x}\right)-\left(\frac{\partial}{\partial z}\Lambda^{\mathbf{1}}\right)\:,\\
0 & = & \delta A_{z}^{\mathbf{2}}=-A(z)\left(\frac{\partial}{\partial z}\xi^{y}\right)-\left(\frac{\partial}{\partial z}\Lambda^{\mathbf{2}}\right)\:,\\
0 & = & \delta A_{z}^{\mathbf{3}}=-\Phi(z)\left(\frac{\partial}{\partial z}\xi^{t}\right)-\left(\frac{\partial}{\partial z}\Lambda^{\mathbf{3}}\right)\:.
\end{eqnarray}
The solutions are 
\begin{eqnarray}
\xi^{t} & = & \tilde{\xi}^{t}(x,y)\:,\\
\xi^{x} & = & -\int dz\frac{1}{r(z)^{2}\sqrt{F(z)}}\left(\frac{\partial}{\partial x}\tilde{\xi}^{z}(x,y)\right)+\tilde{\xi}^{x}(x,y)\:,\\
\xi^{y} & = & -\int dz\frac{1}{r(z)^{2}\sqrt{F(z)}}\left(\frac{\partial}{\partial y}\tilde{\xi}^{z}(x,y)\right)+\tilde{\xi}^{y}(x,y)\:,\\
\xi^{z} & = & \sqrt{F(z)}\tilde{\xi}^{z}(x,y)\:,\\
\Lambda^{\mathbf{1}} & = & \int dz\frac{A(z)}{r(z)^{2}\sqrt{F(z)}}\left(\frac{\partial}{\partial x}\tilde{\xi}^{z}(x,y)\right)+\tilde{\Lambda}^{\mathbf{1}}(x,y)\:,\\
\Lambda^{\mathbf{2}} & = & \int dz\frac{A(z)}{r(z)^{2}\sqrt{F(z)}}\left(\frac{\partial}{\partial y}\tilde{\xi}^{z}(x,y)\right)+\tilde{\Lambda}^{\mathbf{2}}(x,y)\:,\\
\Lambda^{\mathbf{3}} & = & \tilde{\Lambda}^{\mathbf{3}}(x,y)\:,
\end{eqnarray}
where $\tilde{\xi}^{\mu}(x,y)$ is an arbitrary vector function of
$(x,y)$. Then the residual gauge transformations for the vector mode
fluctuations are 
\begin{eqnarray}
\delta g_{tx} & = & F(z)\left(\frac{\partial}{\partial x}\tilde{\xi}^{t}(x,y)\right)\:,\\
\delta A_{x}^{\mathbf{3}} & = & -\Phi(z)\left(\frac{\partial}{\partial x}\tilde{\xi}^{t}(x,y)\right)+\lambda A(z)\int dz\frac{A(z)}{r(z)^{2}\sqrt{F(z)}}\left(\frac{\partial}{\partial y}\tilde{\xi}^{z}(x,y)\right)\nonumber \\
 &  & +\lambda A(z)\tilde{\Lambda}^{\mathbf{2}}(x,y)-\left(\frac{\partial}{\partial x}\tilde{\Lambda}^{\mathbf{3}}(x,y)\right)\:,\\
\delta A_{t}^{\mathbf{1}} & = & -\lambda\Phi(z)\left[\int dz\frac{A(z)}{r(z)^{2}\sqrt{F(z)}}\left(\frac{\partial}{\partial y}\tilde{\xi}^{z}(x,y)\right)+\tilde{\Lambda}^{\mathbf{2}}(x,y)\right]\:,
\end{eqnarray}
and 
\begin{eqnarray}
\delta g_{ty} & = & F(z)\left(\frac{\partial}{\partial y}\tilde{\xi}^{t}(x,y)\right)\:,\\
\delta A_{y}^{\mathbf{3}} & = & -\Phi(z)\left(\frac{\partial}{\partial y}\tilde{\xi}^{t}(x,y)\right)-\lambda A(z)\int dz\frac{A(z)}{r(z)^{2}\sqrt{F(z)}}\left(\frac{\partial}{\partial x}\tilde{\xi}^{z}(x,y)\right)\nonumber \\
 &  & -\lambda A(z)\tilde{\Lambda}^{\mathbf{1}}(x,y)-\left(\frac{\partial}{\partial y}\tilde{\Lambda}^{\mathbf{3}}(x,y)\right)\:,\\
\delta A_{t}^{\mathbf{1}} & = & \lambda\Phi(z)\left[\int dz\frac{A(z)}{r(z)^{2}\sqrt{F(z)}}\left(\frac{\partial}{\partial x}\tilde{\xi}^{z}(x,y)\right)+\tilde{\Lambda}^{\mathbf{1}}(x,y)\right]\:.
\end{eqnarray}
In the above equations, the part that contains $\tilde{\xi}^{t}(x,y)$
is the residual gauge transformation that can be used to obtain the
fourth solution (\ref{FourthSolution}) in the vector mode boundary-to-bulk
propagators.

\bigskip{}

\section{Near-Critical First Order Fields }

Here we give the solutions for $r^{(1)}(z)$, $F^{(1)}(z)$ and $\Phi^{(1)}(z)$.
They are written as indefinite integrals over sources quadratic in
$A^{(1)}(z)$, plus two general solutions to the homogeneous equations
with arbitrary coefficients $C_{1}$ - $C_{6}$, as to be used in
Section 4.7. 
\begin{eqnarray}
r^{(1)}(z) & = & -\frac{1}{2}\int dz\int dz\left[\frac{1}{r^{(0)}(z)}\left(\frac{d}{dz}A^{(1)}(z)\right)^{2}+\frac{\lambda^{2}\Phi^{(0)}(z)^{2}}{r^{(0)}(z)F^{(0)}(z)^{2}}A^{(1)}(z)^{2}\right]\\
 &  & \qquad+C_{1}z+C_{2}\:,\nonumber \\
\Phi^{(1)}(z) & = & -2\int\frac{dz}{r^{(0)}(z)^{2}}\int dz\Bigg\{\left(\frac{d}{dz}\Phi^{(0)}(z)\right)\left[r^{(0)}(z)\left(\frac{d}{dz}r^{(1)}(z)\right)-\left(\frac{d}{dz}r^{(0)}(z)\right)r^{(1)}(z)\right]\nonumber \\
 &  & \qquad-\frac{\lambda^{2}\Phi^{(0)}(z)}{F^{(0)}(z)}A^{(1)}(z)^{2}\Bigg\}+C_{3}\int\frac{dz}{r^{(0)}(z)^{2}}+C_{4}\:,\\
F^{(1)}(z) & = & \int\frac{dz}{r^{(0)}(z)^{2}}\int dz\Bigg\{-2r^{(0)}(z)\frac{d}{dz}\left[F^{(0)}(z)\left(\frac{d}{dz}r^{(1)}(z)\right)\right]\nonumber \\
 &  & \qquad+2\left(\frac{d}{dz}r^{(0)}(z)\right)\left(\frac{d}{dz}F^{(0)}(z)\right)r^{(1)}(z)+r^{(0)}(z)^{2}\left(\frac{d}{dz}\Phi^{(0)}(z)\right)\left(\frac{d}{dz}\Phi^{(1)}(z)\right)\Bigg\}\nonumber \\
 &  & \qquad+C_{5}\int\frac{dz}{r^{(0)}(z)^{2}}+C_{6}\:.
\end{eqnarray}
The trace equation (\ref{EQ5}) at $O\left(\epsilon^{2}\right)$ order
is 
\begin{eqnarray}
\frac{1}{r^{(0)}(z)}\frac{d}{dz}\left[r^{(0)}(z)^{4}\left(\frac{d}{dz}F^{(1)}(z)\right)\right]+4r^{(0)}(z)\frac{d}{dz}\left[r^{(0)}(z)F^{(0)}(z)\left(\frac{d}{dz}r^{(1)}(z)\right)\right]\qquad\\
+2r^{(0)}(z)\left(\frac{d}{dz}r^{(0)}(z)\right)^{2}F^{(1)}(z)-4\left(\frac{d}{dz}r^{(0)}(z)\right)\left[\frac{d}{dz}\left(r^{(0)}(z)F^{(0)}(z)\right)\right]r^{(1)}(z) & = & 0\:.\nonumber 
\end{eqnarray}

\bigskip{}

\end{appendices}

\bigskip{}

\end{document}